\documentclass[10pt, a4paper, onecolumn]{article}

\usepackage{arxiv}

\usepackage[utf8]{inputenc} 
\usepackage[T1]{fontenc}    
\usepackage[breaklinks]{hyperref}
\usepackage{url}
\usepackage{breakurl}
\usepackage{booktabs}       
\usepackage{amsfonts}       
\usepackage{nicefrac}       
\usepackage{microtype}      
\usepackage{amsmath}
\usepackage{amsfonts}
\usepackage{cite}
\usepackage{amssymb}
\usepackage{amsthm}
\usepackage{verbatim}
\usepackage{enumerate}
\usepackage{mdwlist} 
\usepackage{color}
\usepackage[utf8]{inputenc}

\usepackage{algpseudocode}
\usepackage{algorithmicx}
\usepackage{lineno}
\usepackage{framed,multirow}
\usepackage[final]{pdfpages}
\usepackage{latexsym}
\usepackage{wrapfig}

\usepackage{graphicx, subfig}
\usepackage{verbatim}
\usepackage{bm}
\usepackage{authblk}

\expandafter\def\expandafter\UrlBreaks\expandafter{\UrlBreaks
    \do\a\do\b\do\c\do\d\do\e\do\f\do\g\do\h\do\i\do\j%
    \do\k\do\l\do\m\do\n\do\o\do\p\do\q\do\r\do\s\do\t%
    \do\u\do\v\do\w\do\x\do\y\do\z\do\A\do\B\do\C\do\D%
    \do\E\do\F\do\G\do\H\do\I\do\J\do\K\do\L\do\M\do\N%
    \do\O\do\P\do\Q\do\R\do\S\do\T\do\U\do\V\do\W\do\X%
    \do\Y\do\Z\do\/\do-}

\title{Social Interaction Layers in Complex Networks for the Dynamical Epidemic Modeling of COVID-19 in Brazil}

\author[1]{Leonardo F. S. Scabini}
\author[2]{Lucas C. Ribas}
\author[2]{Mariane B. Neiva}
\author[1]{Altamir G. B. Junior}
\author[2]{Alex J. F. Farfán}
\author[1,2]{Odemir M. Bruno}
\affil[1]{\small{S\~{a}o Carlos Institute of Physics, University of S\~{a}o Paulo (USP), PO Box 369, 13560-970, S\~{a}o Carlos, SP, Brazil. \protect\\Scientific Computing Group}}
\affil[2]{\small{Institute of Mathematics and Computer Science, University of S\~{a}o Paulo (USP), USP, Avenida Trabalhador s\~ao-carlense, 400, 13566-590, S\~ao Carlos, SP, Brazil.}}

\begin{document}
\maketitle

\begin{abstract}
We are currently living in a state of uncertainty due to the pandemic caused by the Sars-CoV-2 virus. There are several factors involved in the epidemic spreading such as the individual characteristics of each city/country. The true shape of the epidemic dynamics is a large, complex system such as most of the social systems. In this context, Complex networks are a great candidate to analyze these systems due to their ability to tackle structural and dynamical properties. Therefore this study presents a new approach to model the COVID-19 epidemic using a multi-layer complex network, where nodes represent people, edges are social contacts, and layers represent different social activities. The model improves the traditional SIR and it is applied to study the Brazilian epidemic by analyzing possible future actions and their consequences. The network is characterized using statistics of infection, death, and hospitalization time. To simulate isolation, social distancing, or precautionary measures we remove layers and/or reduce the intensity of social contacts. Results show that even taking various optimistic assumptions, the current isolation levels in Brazil still may lead to a critical scenario for the healthcare system and a considerable death toll (average of 149,000). If all activities return to normal, the epidemic growth may suffer a steep increase, and the demand for ICU beds may surpass 3 times the country's capacity. This would surely lead to a catastrophic scenario, as our estimation reaches an average of 212,000 deaths even considering that all cases are effectively treated. The increase of isolation (up to a lockdown) shows to be the best option to keep the situation under the healthcare system capacity, aside from ensuring a faster decrease of new case occurrences (months of difference), and a significantly smaller death toll (average of 87,000).
\end{abstract}


\section{Introduction}

Although we have experienced several pandemics throughout history, COVID-19 is the first major pandemic in the Modern Era. The last critical global epidemic occurred in 1918 and became known as the Spanish flu. But, in 1918, the reality was quite different. Scientific and medical knowledge was much more limited, making it difficult to fight the disease. Furthermore, the world was not globalized, the means of transport were not as agile as the current ones and the population was much smaller. The 21st century is marked by globalization and an intricate and intense social network, which connects in one way or another to everyone on the planet. The latter fact increases the danger that a local epidemic disease will rapidly evolve into a pandemic like what happened in Wuhan, China, and now is all over the world. 

The form of propagation and contagion of the Sars-CoV-2 virus occurs by direct contact between individuals, through secretions, saliva, and especially by droplets expelled during breathing, speeching, coughing, or sneezing. The virus also spreads by indirect contact, when such secretions reach surfaces, food, and objects \cite{WHO2020}. Besides, infected people take a few days to manifest symptoms, which can be severe or as mild as a simple cold. There is even a large proportion of infected people who remain asymptomatic \cite{KIM2020}. This makes it practically impossible to quickly identify the infected and apply effective measures to limit the spread of the disease. Also, Sars-CoV-2 was discovered in December 2019, which makes it very recently in the face of the current epidemic. Little is known about the COVID-19 disease, which appears to be highly lethal, with no drugs to prevent or treat. The concern is greater since direct (individual - individual) and indirect (individual - objects - individual) social relations are the means of spreading the disease. Thus, the social interaction structure is the key to create strategies and guide health organizations and governments to take appropriate actions to combat the disease. 

\begin{figure}[!htb]
    \centering
   \includegraphics[width=0.9\linewidth]{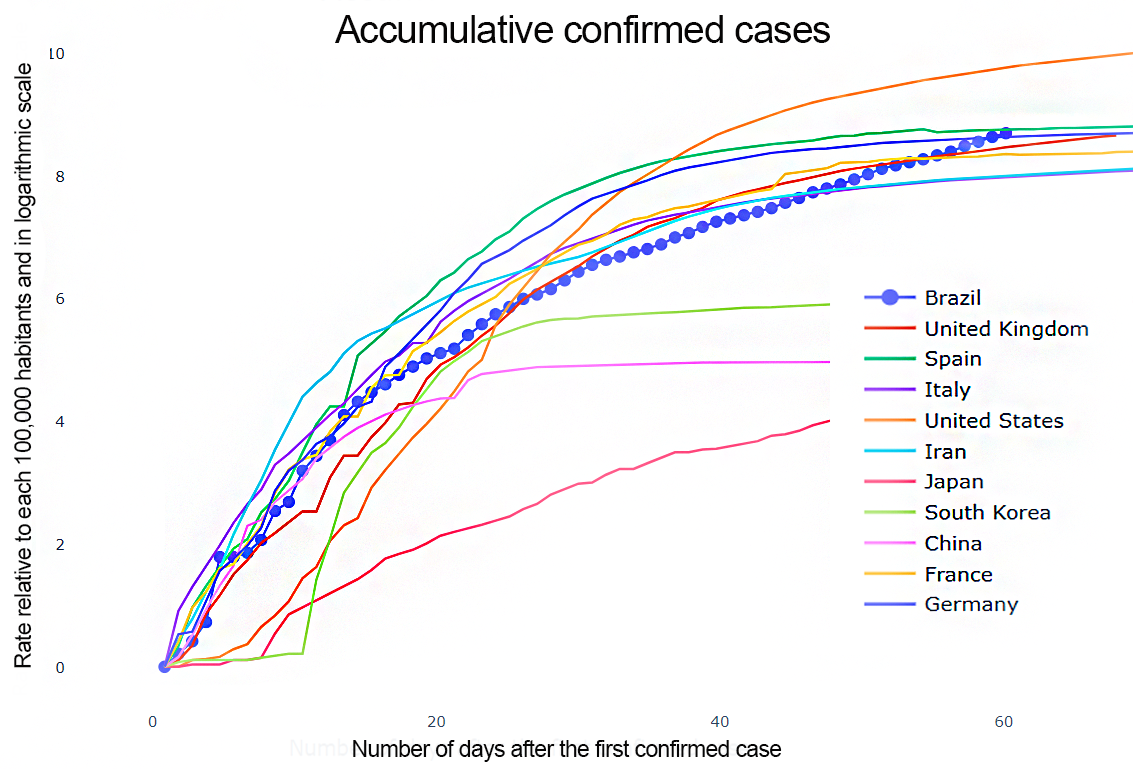} \  
        
    \caption{\label{fig:compared} Total number of cases reported in Brazil compared to other countries (May 5, 2020 \cite{johnsHopkins}). It is possible to notice that Brazil is surpassing countries such as Italy, South Korea, Japan, and China, and it is reaching the relative number of cases in the United Kingdom and France. As of the date of this study, the United States is the epicenter of the pandemic.}
\end{figure}

One of the main concerns is overloading the health system. The first case in Brazil was confirmed on February 26, a 61-year-old man who traveled to the Lombardy region in northern Italy. Now, in the middle of May, there are more than 200,000 cases and 14,000 deaths in all states of Brazil \cite{covid2020brasil}. The concern is even worse due to the country's social inequality, over 80\% of the population relies solely on the public health system and this distribution is not uniform. 
According to \cite{castro2020demand}, there are only 9 hospital beds per 100,000 people in the North region while Southeast accounts for 21 hospital beds. The treatment of severe cases requires the use of respirators/ventilation in intensive care units (ICU), and if simultaneous infections occur there will be no beds to meet the demand and a possibly large number of victims. Thus, it is urgent to develop models and analyses to try to predict the evolution of the virus. Also, as noted in Figure \ref{fig:compared}, Brazil is running towards being the next epicenter of the pandemic. It has already exceeded the number of cases in important countries such as Germany, China, Japan, Italy, Iran, South Korea, and France (the rates consider the population size of each country and are on a logarithmic scale).

Since COVID-19 presents a unique and unprecedented situation, this work proposes a specific model for the current pandemic. Based on the classic epidemic model SIR, also extended to SID\cite{qi2020model}, SIASD\cite{bastos2020modeling} and SIQR\cite{crokidakis2020data}, we propose a more realistic model to better represent the effects of the COVID-19 disease by adding more infection states. The proposed approach also considers social structures and demographic data for complex network modeling. Each individual is represented as a node and edges represent social interaction between them. The multi-layer structure is implemented by different edges representing specific social activities: home, work, transports, schools, religious activities, and random contacts. The probability of contagion is composed of a dynamic term, which depends on the circumstances of the social activity considered, and a global scaling factor $\beta$ for controlling characteristics such as isolation, preventive measures, and social distancing. 

The proposed model can be used to analyze any society given sufficient demographic data, such as medium/big cities, countries, or regions. Here we analyze in depth the Brazilian data. The SIR model is applied through the network using an agent model, and each iteration of the system is simulated using the 24-hour pattern, allowing us to understand the dynamics of the disease throughout the days. The results show the importance of social distancing recommendations to flatten the curve of infected people over time. This is currently maybe the only way to avoid a collapse of the health system in the country.

The paper is divided as follows: Section 2 presents important concepts about complex networks, the SIR model and its applications. Section 3 explains our proposed approach and Sections 4 and 5 presents the results, discussion, and conclusions of the work.

\section{Epidemic Propagation on Complex Networks}

Created from a mixture of graph theory, physics, and statistics, Complex Networks (CN) are capable to analyze not only the elements themselves but also their environment to find patterns and obtain information about the dynamics of a system. As most of the natural structures are composed of connected elements, graphs are suitable to analyze most of the real-world phenomena. Over the past two decades researchers have been showing that many real networks do not present a random structure, and its emergent patterns can be used to understand and characterize a model \cite{barabasi1999emergence, watts1998collective}. Complex network analysis has then been applied to sociology, physics, nanotechnology, neuroscience, biology, among other areas \cite{costa2008complex,costa2011analyzing}. 

To start with a formal definition, a graph G is a set  \{V,E\} where V is composed by N vertices (also known as nodes or elements) $\{v_i, v_n\}$ and E is the set  ${e(v_i,v_j)}$ of edges (or connections) among its elements. Edges represents the relationships between two elements and its value can also represent the strength or weight of a connection if $e(v_i,v_j) > 0$. For a G with only $e(v_i,v_j) = \{0,1 |\text{ } \forall_{i,j}$ $0 < i,j \leq n\}$ the model is an unweighted graph. Furthermore, a network is undirected if $e(v_i,v_j) = e(v_j,v_i)$ and directed, otherwise. 

Usually, applications with complex networks consist of two main steps: i) transform the real structure into a complex network, and ii) analyze the model and extract its features or understand its dynamics. One natural phenomenon that has a straight forward connection to a complex network in society. People are connected due to several aspects such as members of a family, religious groups, co-workers, members of the same school, or faculty, among other social relationships. Therefore CNs have been widely employed for social network analysis \cite{vega2007complex}.

Extended from social interactions, the epidemic spread has also been studied by researchers in the last decades. In this context, one of the best known and widely used epidemic models in infectious diseases is the susceptible-infected-recovered (SIR) model, which is composed of three categories of individuals \cite{bailey1975mathematical,anderson1992infectious}

\begin{itemize}
\item \textbf{Susceptible:} the ones who are not infected but could change its status to a state to infected if in contact with a sick person combined with a probability $\beta$ of contagion
\item  \textbf{Infected:} the ones that have the disease
\item \textbf{Recovered:} usually after some time, a person recovers from the illness and it is not able to be infected again due to the immunity process (in this case, this is an assumption of the process). The recovery rate of infected people is aligned with a probability of $\gamma$
\end{itemize}

Also, the model can be described as

\begin{equation}
\frac{ds}{dt} = -\beta is, \text{   }  \frac{di}{dt} = \beta is - \gamma i, \text{    } \frac{dr}{dt} = -\gamma i
\end{equation}\label{eq:ode}
where \textit{s}, \textit{i} and \textit{r} represents the ratio of susceptible, infected and recovered people in the population, respectively. Usually, the problem is solved with differential equations, however, agent-based techniques in networks can represent the nature of the spread of viral diseases in a more complex scenario. 

If a network is fully connected, meaning that $e(v_i,v_j)$ = $\{1, \forall_{i,j} 0 < i,j <= N\}$, Equation \ref{eq:ode} fits the structure perfectly. However, in the real world, not everyone is connected and people only contract the disease if in contact with an infected individual or object. This is why a complex network approximates the dynamics of real viruses
and can help us to understand the disease behavior. There are various approaches to represent people and society as networks, named social network analysis. Small world networks \cite{moore2000epidemics} can be used as a good approximation of the social connections. In 2000, Moore \cite{moore2000epidemics} emphasized that the use of small-world networks, where the distance among two elements is usually small in comparison to the size of the population, showed a faster spread of the viral disease than classical diffusion methods. The approximation of real social phenomena was first explained by Milgram \cite{milgram1967small} in \cite{milgram1967small}, the sociologist is the author of the well-known idea that there are up to six people separating any two individuals in the world, which reinforces the importance of analyzing the epidemic spread from a graph view. In \cite{newman1999scaling}, the authors used small-world networks to simulate a SIR model, however, they considered that every contact with an infected person resulted in contamination, which is not realistic. Therefore, other researchers improved the model over the years, adding new constraints to approximate the simulation to real scenarios \cite{da2019epidemic}. 

The SIR model on networks works as follows: each node represents a person and, the elements are connected according to some criteria and the epidemic propagation happens through an agent-based approach. It starts from a random node, and for each time step nodes with the susceptible state can contract the disease from a linked infected node with a predefined probability. The same idea occurs with the recovered category. After a certain period, a node can recover or can be removed from the system (case of death) according to a certain probability. At the end of the evolution of a SIR model applied to a network, the number of nodes in each SIR category (susceptible, infected and recovered) can be calculated for each unit of time evaluated and then compare these data with real information, for example, the hospital capabilities of the health system. Also, the probability of infection and recovery can be adjusted over time considering social distancing, hygiene, and health conditions. 

\section{Proposed Model:  COmplexVID-19} 

The proposed model extends the SIR model to a more realistic scenario to achieve a better correlation to the COVID-19 disease, since the model was created specifically for the disease, we named the model as COmplexVID-19. Our strategy is based on a multi-layer network to represent the Brazilian demography and its different characteristics of social relationships. Each layer is composed of a set of groups representing how people interact in a given social context. In the network, a node represents a person and the edges are the social relationships between persons, and they are also the means through which the disease can be transmitted. The virus spreads from an infected node to neighboring nodes at each iteration step (1 step = 1 day), according to a given infection probability. First, we describe how the layers are built based on social data from Brazil.

\subsection{Network Layer Structure Over Brazilian Demography}

To define the different social relations, the first information needed is the age distribution so that groups such as schools and work can be separated. We consider the Brazilian age distribution in relation to the total population in 2019 \cite{ibgeages}, details are given on Table \ref{table:brazildemog}. This distribution is used to define an age group for each node, which is then used to determine its social activities through the creation of edges on different layers. In this approach, each network-layer represents a kind of social relationship or activity that influences the transmission of the COVID-19. In this way, it is possible to evaluate and understand what is the impact of each social activity in the epidemic propagation. Basically, in this work, a network layer is represented by a set of edges connecting some nodes. The following social activities are considered, composing 6 different layers:
\begin{itemize}

    \item \textbf{Home:} in this layer, all people that live in the same residence are connected.
    
    \item \textbf{Work:} connects people that work in the same environment/company.

    \item \textbf{Transport:} this layer represents people that eventually take the same vehicle at public transports.
  
    \item \textbf{School:} represents the social contact of students that belong to the same school class.

    \item \textbf{Religious activities:} connects people of the same group of some religious activity.
   
   \item \textbf{Random:} this layer represents activities of smaller intensity, such as indirect contact (through objects/surfaces).

\end{itemize}

\begin{table}[!htb]   
    \centering
    \caption{Demographic data for Brazil. Source: IBGE 2010  \cite{ibgefamily} and 2019 \cite{ibgeages} census. \label{table:brazildemog} }               
     
    \begin{tabular}{|c|c|c|c|c|c|c|}  
    \hline
       \textbf{Age distribution} & \textbf{Quantity} &  \textbf{Fraction ($\%$)}  \\
        \hline\hline
        
        0-13    & 38,464,000    & 18\\
        14-17    & 12,518,000    & 6\\
        18-24    & 22,068,000    & 11\\
        25-39    & 47,577,000    & 23\\
        40-59    & 55,455,000    & 26\\
        60+        & 33,995,000    & 16\\
        \hline
        Total population (2019)    & 210,077,000 & 100\\    

     \hline
    \end{tabular}  
    \begin{tabular}{|c|c|c|c|c|c|}  
    \hline
       \textbf{Family size} & \textbf{Fraction ($\%$)} \\
        \hline\hline

       $1$ person & $12$\\\hline
       
       $2$ persons & $22$\\\hline
       
       $3$ persons & $25$\\\hline
       
       $4$ persons & $21$\\\hline
       
       $5$ persons & $11$\\\hline
       
       $6$ persons & $5$\\\hline
       
       $7$ persons & $2$\\\hline

       $8$ persons & $1$\\\hline
       $9$ persons & $0.5$\\\hline
       $10$ persons & $0.5$\\\hline

    \end{tabular}  
    
\end{table}   

\begin{figure}[!htb]
    \centering
    
    \subfloat[Connections of layer "home".]{ \includegraphics[width=0.32 \linewidth]{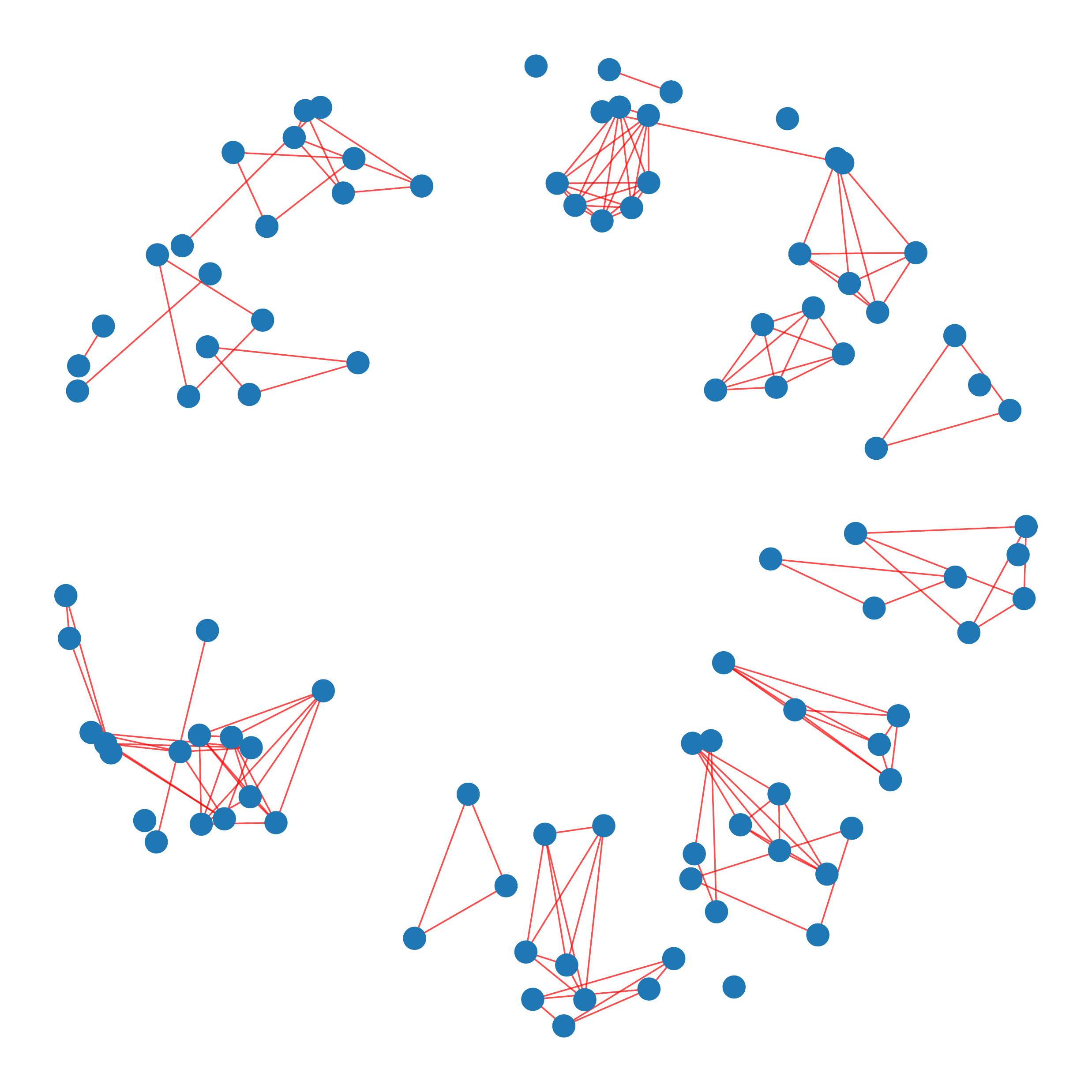}}  \subfloat[Connections of layer "work".]{ \includegraphics[width=0.32 \linewidth]{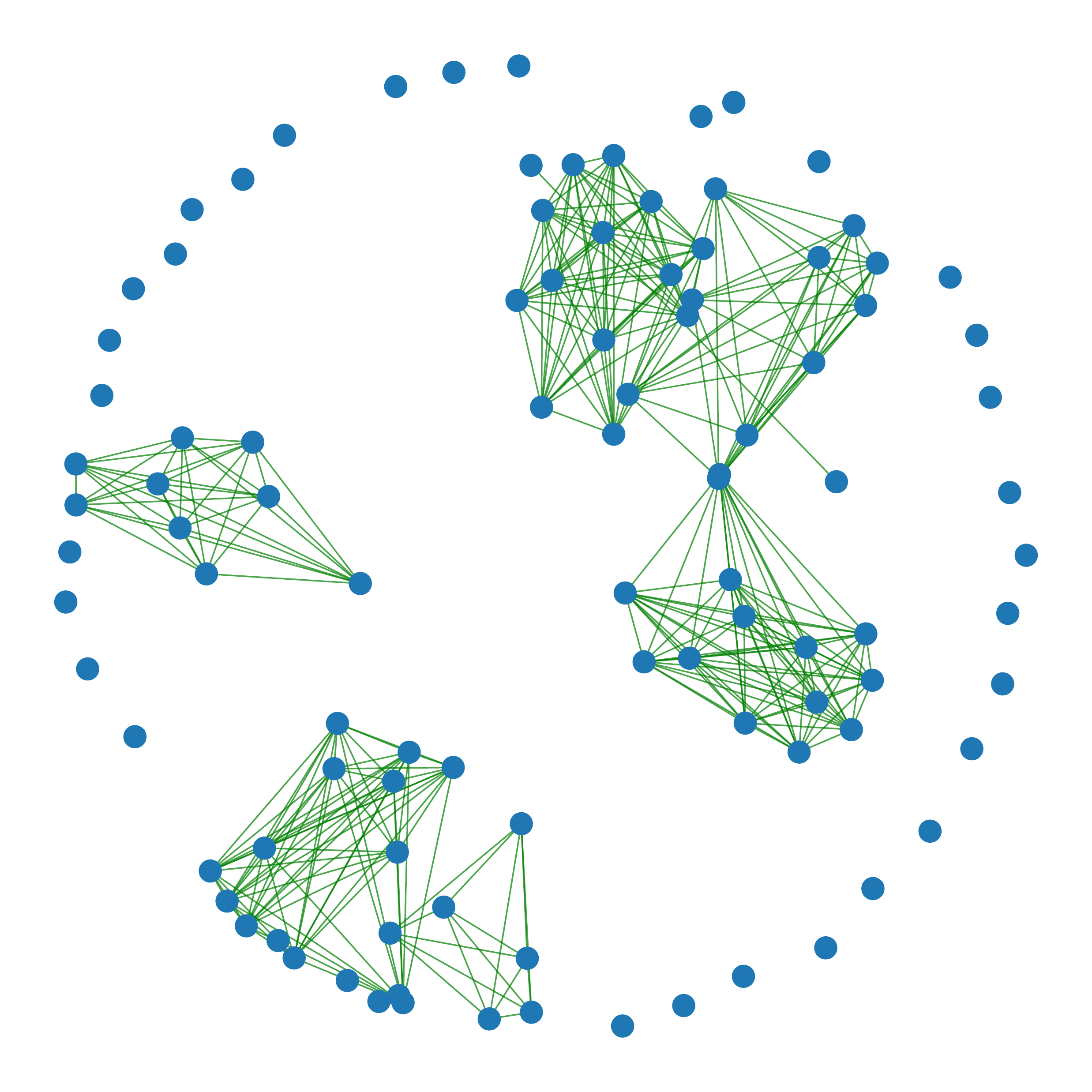}}  \subfloat[Connections of layer "transport".]{ \includegraphics[width=0.32 \linewidth]{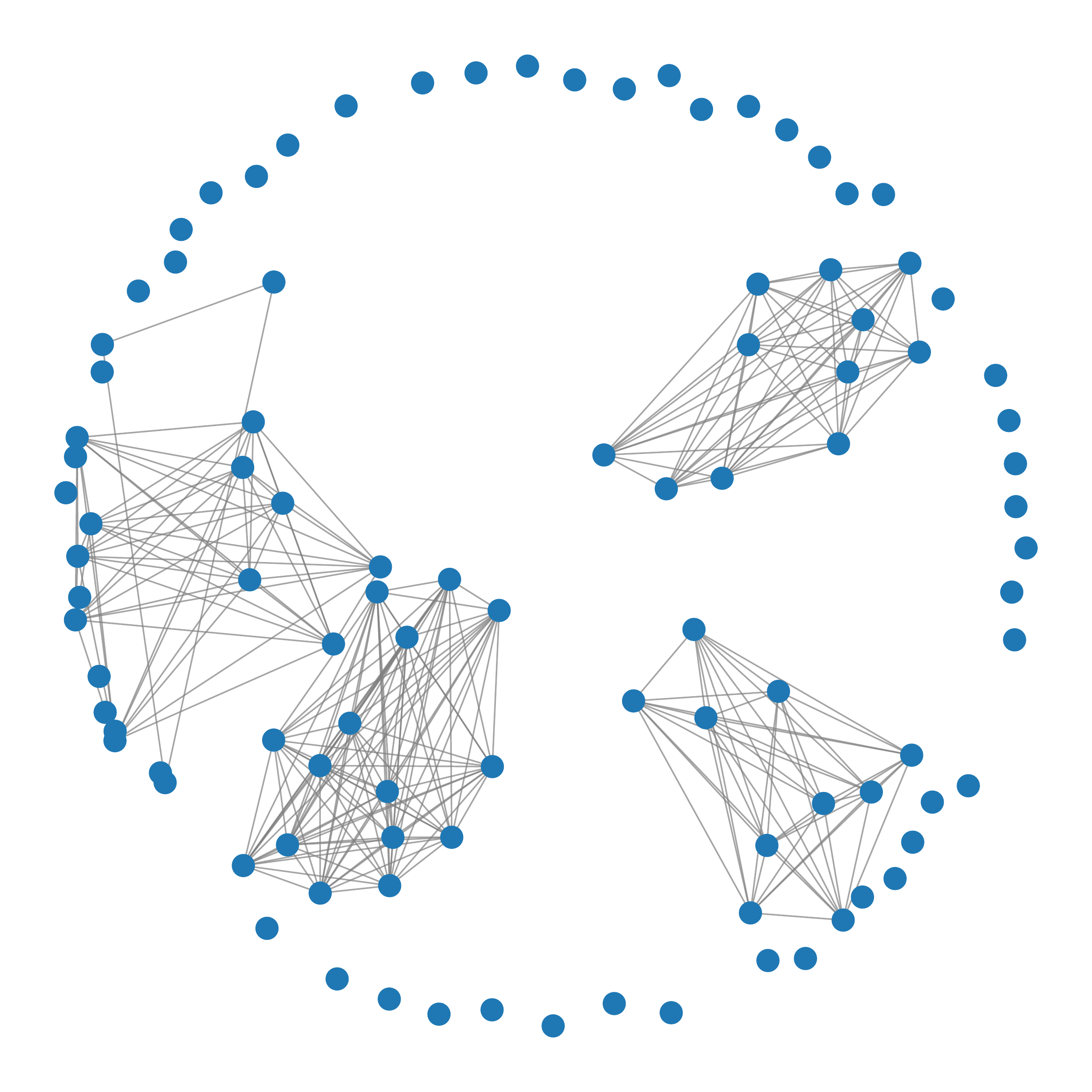}}
    \\
    \subfloat[Connections of layer "school".]{ \includegraphics[width=0.32 \linewidth]{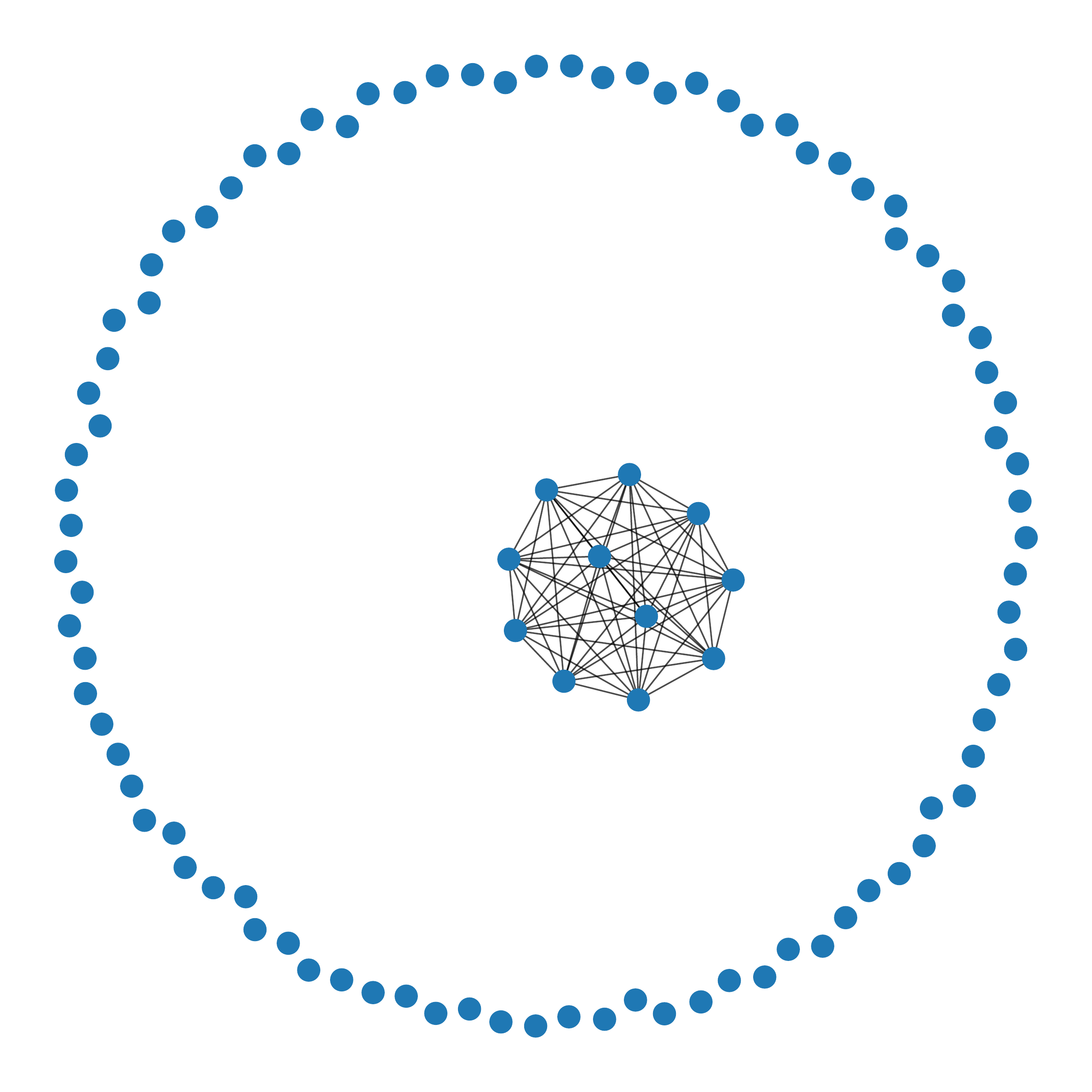}}  \subfloat[Connections of layer "religion".]{ \includegraphics[width=0.32 \linewidth]{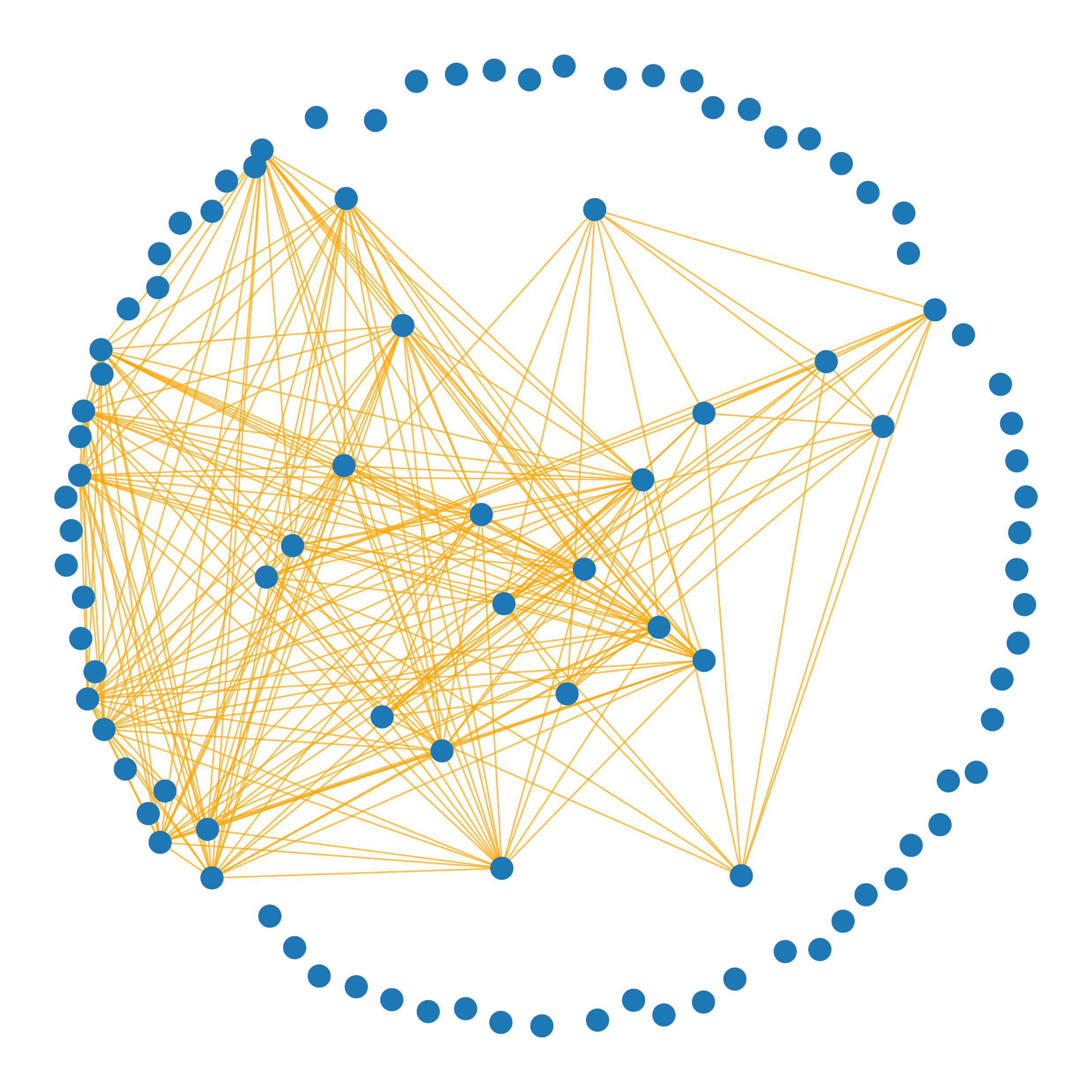}}  \subfloat[Connections of layer "random".]{ \includegraphics[width=0.32 \linewidth]{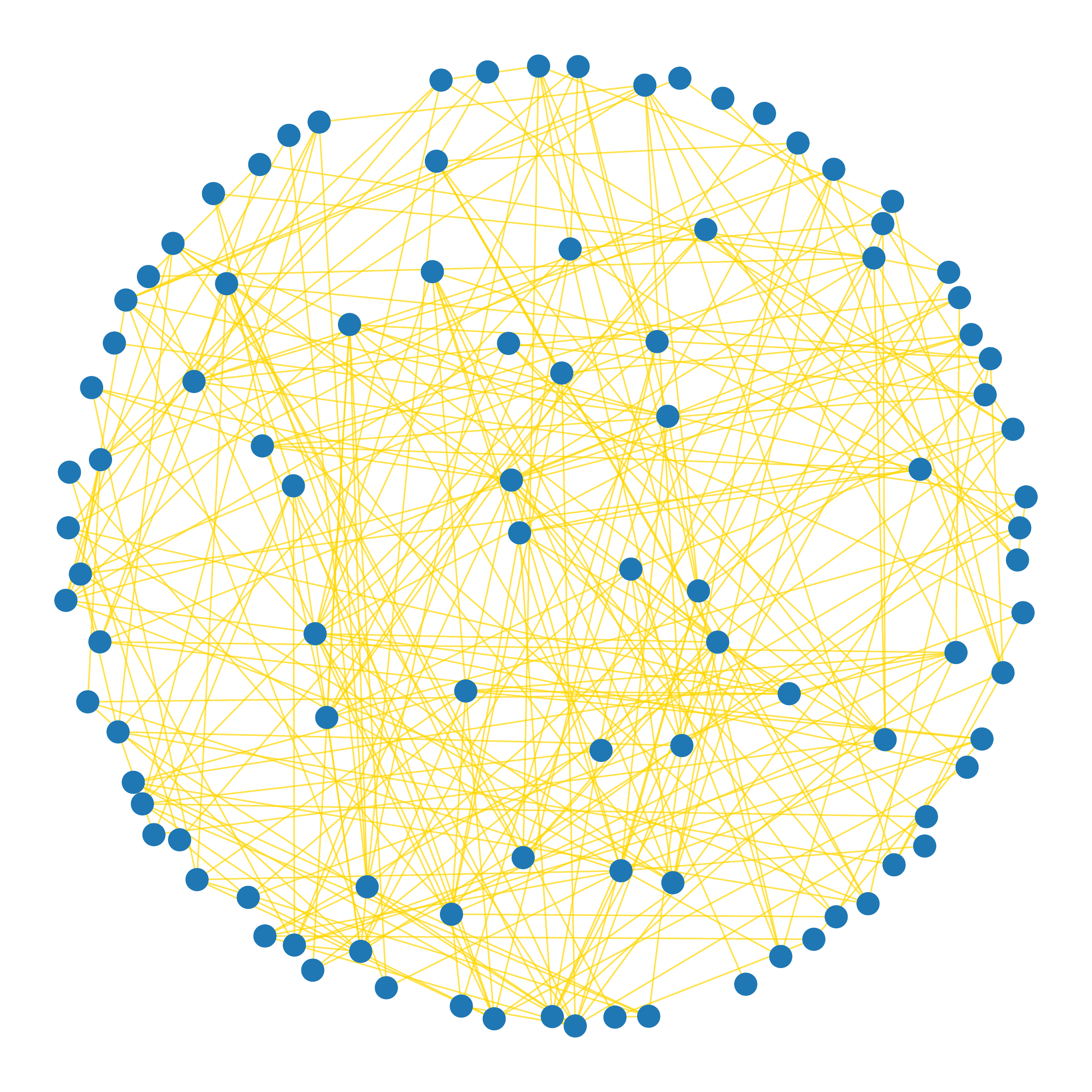}}
    \\
        
    \caption{\label{fig:networklayers} Each social layer of the proposed multi-layer network. The nodes are people and do not change across layers, and the weighted connections represent social contact which may lead to infection according to the edge weight (probability value between $[0,1$).}
    
\end{figure}

The first layer represents home interactions and is composed of a set of groups with varying size which are fully connected internally. These groups have no external connections, i.e. the network starts with disconnected components representing each family. To create each group, we consider the Brazilian family size distribution for 2010 \cite{ibgefamily}, the year with more detailed information on family sizes from 1 up to 14 members. We consider the probability of a family having sizes from 1 to 10, therefore the probability of a family having 10 persons is the sum of the higher sizes, the details of this distribution are given in Table \ref{table:brazildemog}. The first layer is then created following the family size distribution and ensuring that each family has at least 1 adult. Figure \ref{fig:networklayers} (a) shows the structure of such a layer built for a population of $n=100$.

A large fraction of the population in any country needs to work or practice some kind of economic activity, which also means interacting with other people. Thus, work represents one of the most important factors of social relations, which is also very important in an epidemic scenario. To represent the work activity we propose a generic layer to connect people with ages from 18 to 59 years, i.e. $60\%$ of the total population in the case of Brazil. There is a wide variety of jobs and companies, therefore it is not trivial to create a connection rule that precisely reflects the real world. Here, we consider an average scenario with random groups of sizes around $[5,30]$, uniformly distributed, and internally connected (such as the  "home" layer). An example of this layer is shown on Figure \ref{fig:networklayers} (b), using $n=100$. Although the nodes of a group are fully connected, the transmission of the virus depends directly on the edge weights, which we discuss in-depth on Section \ref{sec:infection}.

Collective transports are essential in most cities, however, it is one of the most crowded environments and plays an important role in an epidemic scenario also due to the possibility of geographical spread, as vehicles are constantly moving around. The third layer we propose represents this kind of transports, such as public transports, and includes people that do not possess or use a personal vehicle. In Brazil the number of people using public transport depends on the size of the city, with $64.98\%$ in the capitals and $35.89\%$ in other cities \cite{transportsIPEA}, with an average use of around 1.2 hours a day \footnote{\url{http://g1.globo.com/bom-dia-brasil/noticia/2015/02/brasileiros-gastam-em-media-1h20-por-dia-em-transportes-publicos.html} \label{f1}}. Here we consider the average of the population between the two cases ($50\%$), randomly sampled, to participate in the "transports" layer. Random groups are created with sizes between $[10,40]$, uniformly sampled, and the nodes within each group are fully connected. This variation of sizes is considered to represent cases such as low and high commuting times, and also the differences between vehicle sizes. Other factors such as agglomeration and contact intensity are discussed in Section \ref{sec:infection}. This layer is illustrated on Figure \ref{fig:networklayers} (c).

Schools are another environment of great risk for epidemic propagation. The proposed layer considers the characteristics of schools from primary to high school and how children interact. We consider that all persons from 0 to 17 years ($24\%$ of the Brazilian population) participate in this layer, and the size of the groups, which represents different school classes, varies uniformly between $[16,30]$ \cite{INEP}. This layer is illustrated on Figure \ref{fig:networklayers} (d).

Brazil is a very religious country, in which by 2010 only around $16.2\%$ of the population claimed not to belong to any religion \cite{ibgereligion}. $64.6\%$ claimed to be catholic and $22.2\%$ to be protestant, summing up to $86.8\%$ of the total population. Here we consider that nearly half of these people ($40\%$ of the total population) actively participate in religious activities (weekly). The distribution of religious temple sizes is defined as a Pareto distribution in the interval $[10,100]$. Taking into account that wage distribution follows the Pareto distribution approximately, we model real estate predominance according to their capacity. The assumption here is that building costs (for churches, offices, homes, etc.) have a linear relationship to their internal capacity, and thus any given capacity has a power-law relationship with the number of such buildings within a region.

We consider a random layer to represent all kinds of contacts not related to the specific previous social layers. This includes small direct contacts (person-to-person) and indirect contacts (individual - objects - individual) that may happen throughout the week, such as random friend/neighbor meetings, shopping, and other activities that involve surface contacts. For that $5n$ new random edges are created, that can connect any node. On the one hand, this yields an average of 5 random connections to each node, which can randomly connect any other node. On the other hand, the impact of this layer on the epidemic is smaller than the others, as it represents rapid contacts in comparison to the other activities described, thus its infection probability is smaller. In the following section we discuss the details concerning this aspect, deriving from the edge weights of each layer. In Figure \ref{fig:networklayers} (f) an example of this layer is shown. The overall structure of social interactions in our model can be compared to the statistical analysis in \cite{ferguson2020report}, however here we introduce a more detailed model of social contacts with specific layers and connection patterns to better fit the particularities of a given country or city.

\subsubsection{Infection probabilities}\label{sec:infection}

Unlike the traditional SIR model, which consists of a single $\beta$ term to describe the probability of infection, here we propose a dynamic strategy to better represent the real world and the new COVID-19 disease. The idea is to incorporate important characteristics in the context of epidemic propagation according to each layer. Firstly, to a given layer a fixed probability term is calculated to represent its characteristic of social interaction. For this, we considered 3 local terms: the contact time per week, the average number of people close to each other (agglomeration level), and the total number of people involved in the respective activity. Considering two nodes $v_x$ and $v_y$, connected at group $j$ of layer $i$, its edge weight is then defined by
\begin{equation}
    e(v_x, v_y) = \frac{t_i}{168} \frac{k_i}{n_{ij}}  \beta
\end{equation}
where $t_i$ represents the average weekly contact time on layer $i$, $k_i$ is the agglomeration level (average number of nearby people) and $n_{ij}$ represents the size of the group $j$ in which the nodes participates on layer $i$. The first fraction represents the contact time normalized by the total time of the week ($24*7=168$), and the second fraction represents the proportion among the local people closest to the total number of people on that activity group.

The first part of the infection probability equation is multiplied by a $\beta$ term, which scales the original probability. The $\beta$ term is then the only parameter to tune the infection rates for the entire network, and the other properties are specific for the studied society, based on its population characteristics and the nature of the activities (layers). Table \ref{table:averagepf} shows these specific properties that we considered for the Brazilian population, and how the infection probabilities are calculated for each layer. In the table, we have the following information: who or how many people are part of the activity represented by a layer (column "who", discussed in the previous section); contact time according to activity (column "Time of contact"); the average number of people close to each other in each activity (column "Nearest", represents the agglomeration level); the number of connections between people (column "Group size"); the probability of infection (column "Probability").

    \begin{table}[!htb]   
    \centering
    \caption{Specific brazilian properties considered to compose each social layer and calculate their probability of infection, i.e. the edge weights of each layer.}                                                                                                   
    \label{table:averagepf}  
     \resizebox{\linewidth}{!}{
    \begin{tabular}{|c|c|c|c|c|c|}  
    \hline
       \textbf{Activity} & \textbf{Who} & \textbf{Time ($t_i$)} & \textbf{Nearest ($k_i$)} & \textbf{Group size ($n_{i,j}$)} & \textbf{Infection Prob. ($e$)}\\
        \hline
         \hline
       Home & everyone & 3 hours a day& $k_i = n_{i,j}$ & $[1,10]$, Table \ref{table:brazildemog} & $(\frac{21}{168}  \frac{n_{i,j}}{n_{i,j}}) \beta$\\
        \hline
       Work & 18 to 59 years& 8 hours a day, 5 days & 3  & $[5,30]$, uniform & $(\frac{40}{168}  \frac{3}{n_{i,j}}) \beta$\\
        \hline
       Transports & $50\%$, random& 1.2 hours a day & 8  & $[10,40]$, uniform & $(\frac{8.4}{168}  \frac{8}{n_{i,j}}) \beta$\\
        \hline
       Schools & 0 to 17 years& 4 hours a day, 5 days& 5  & $[16,30]$, uniform & $ (\frac{20}{168}  \frac{5}{n_{i,j}}) \beta$\\
        \hline
       Religion & $40\%$, random& 2 hours a week & 6  & $[10,100]$, Pareto & $ (\frac{2}{168}  \frac{6}{n_{i,j}}) \beta$\\
        \hline
       Random & 5 per person & 1 hour a week & 1  & 1-to-1 contacts & $(\frac{1}{168}) \beta$\\
       
    \hline

     \hline
    \end{tabular}  
    }
\end{table}

\subsection{Dynamics Modeling}

    The proposed model is a variant of the SIR approach where we include new possible states, structural and dynamic mechanisms after the new findings on COVID-19. The traditional SIR model consists of 3 states: Susceptible, Infected, and Recovered. To better represent the intrinsic dynamics of the new epidemic, we considered 7 states according to reported distributions of the clinical spectrum \cite{lauer2020incubation,surveillances2020epidemiological}:
    
    \begin{itemize}
        \item \textbf{Susceptible}: Traditional case, it means that a person can be infected at any time. This is the initial state of every node.
        
        \item \textbf{Infected - asymptomatic}: People who do not show any symptoms ($30\%$ of the total cases of infection) and remain contagious for up to 18 days (they may recover after 8 days). This is the most dangerous case for the epidemic spreading because the person is not aware of its infection.
        
        \item \textbf{Infected - Mild}: $55\%$ of the cases, present mild and moderated symptoms with no need for hospitalization, remain contagious for up to 20 days, and may recover after 10 days of infection.
        
        \item \textbf{Infected - Severe}: $10\%$ of the cases, present strong symptoms, and need hospitalization, remain contagious for up to 25 days. Has a death rate of $15\%$ and may recover after 20 days.
        
        \item \textbf{Infected - Critical}: Present worst symptoms and remain contagious for up to 25 days, need ICU and Ventilation, have a death rate of $50\%$ and may recover after 21 days.
        
        \item \textbf{Recovered}: People who went through one of the infection cases and overcame the disease, ceasing to contaminate and supposedly becoming immune. These nodes no longer interact with other nodes anymore and are therefore removed from the network.
        
        \item \textbf{Dead}: People who went through severe or critical cases and eventually died. These nodes are also removed from the network.

    \end{itemize}
    
    Estimates for the proportion of asymptomatic cases vary from $18\%$ ($95\%$ confidence, $[15.5,20.2\%$]) \cite{mizumoto2020estimating} to $34\%$ ($95\%$ confidence, $[8.3,58.3\%]$) \cite{huang2020clinical}. Considering the confidence intervals, here we roughly approximate it to an average of $30\%$ of the total number of infected cases. However, it is very difficult to study asymptomatic cases due to several reasons, such as the lack of available tests and the difficulty in identifying potential cases, which would include every person who had contact with known symptomatic cases. Some studies indicate that asymptomatic cases may remain contagious for up to 25 days, with an incubation period of 19 days \cite{bai2020presumed}, but the viral load may be smaller at the end of the infection. Here we take an optimistic approach considering that they may recover (become immune and cease to contaminate) uniformly after 8 days of infection, up to around 18 days. As for the recovered nodes, we are considering that people become immune or at least acquire a long-term resistance to the virus, up to a maximum of 300 days (limit of our simulations). However, this should be taken cautiously as these properties are not yet fully understood \cite{lan2020positive}.

    \subsubsection{Dynamic Evolution}

    The infection grows through the contact (edges) between infected and susceptible nodes, and the probability of being infected is the edge weight. If infection occurs, then one of the 4 infection cases are chosen based on the probability described above ($30\%$, $55\%$, $10\%$ and $5\%$). This distribution plays an important role in the structure and dynamics of the network. The node structure of asymptomatic cases does not change during the simulation, except for the time it takes to cease contamination and recover. It means that as these persons are not aware of their contamination, they will remain acting normally on the network (according to the active layers and edge weights). Their contagious time varies from 1 to 18 days after infection.

    Concerning the other cases (mild, severe, and critical), we consider the incubation time of the virus, the recovery time, the contagion time, the death rates of each case, and the usual action taken by the infected person or health professionals at hospitals. Various works \cite{linton2020incubation,backer2020incubation,lauer2020incubation} point out that the average incubation period of COVID-19 is around 5 days, but some cases may take much less or more time. The official WHO report \cite{world2020coronavirus} states that the average incubation time is around 5 to 6 days, with cases up to 14 days. The results in \cite{lauer2020incubation} show that the average shape of the incubation time follows a log-normal distribution (Weibull distribution) with an average of $6.4$ days and a standard deviation of $2.3$ days. In this context, we consider the day when an infected person begins to show symptoms by randomly sampling from this distribution (1000 repetitions), with cases varying from 2 to 14 days.

\begin{figure}[!htb]
        \centering
       \includegraphics[width=1 \linewidth]{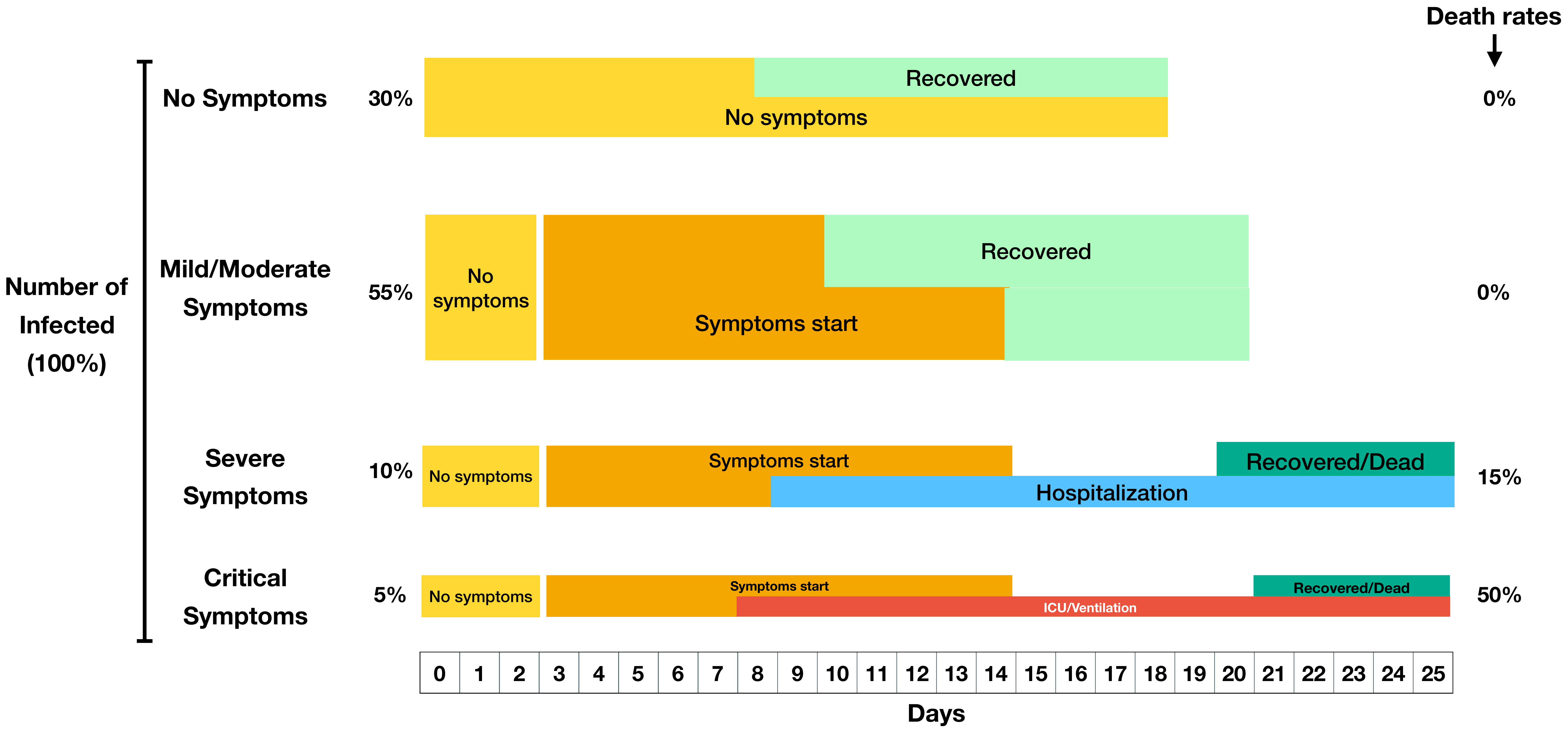} 
        \caption{\label{fig:dynamic_configuration} 
        Configuration considered for the dynamic evolution of each type of infected node in the proposed SIR model. Each overlapping region is treated as a combined probability distribution that defines when one phase ends for the other to begin.}
    \end{figure}

    For mild cases, the nodes are isolated at home, maintaining the connections of the first layer, and then only $20\%$ of the cases are diagnosed. Considering the ratio of diagnosed cases, patients who are asymptomatic or with mild symptoms of COVID-19 may not seek health care, which leads to the underestimation of the burden of COVID-19 \cite{lai2020asymptomatic}. Moreover, our diagnosis rule is also based on the fact that ongoing tests in Brazil are increasing more slowly than in most European countries and the USA (tests are being performed mostly on people that need hospitalization). If a given case is severe or critical, the patient goes to a hospital and is fully isolated, i.e. we remove all of its connections. This is a rather optimistic assumption, considering that these patients still may infect the hospital staff. Concerning the time that patients usually stay at hospitalization/ICU, the works \cite{bhatraju2020covid,zhou2020clinical} points to an average of 14 days for all cases. For standard hospitalization, we considered a minimum of 6 days and a maximum of 16 days of stay, and for the ICU/Ventilation, a minimum of 7 and a maximum of 17 days of stay. The time of each case will depend on the day the symptoms start and the day of recovering/death. Figure \ref{fig:dynamic_configuration} illustrates all the infected states and mechanisms described here. This configuration results in an overall lethality of $4\%$. It is important to stress that here we consider a maximum of 25 days of infection time, which is the time frame based on most studies we have seen so far in the literature. We are still at the beginning of the pandemic and a better characterization of the long-term impact is very difficult. Nonetheless, the available information allows to represent the most obvious features of the Sars-CoV-2 virus and to evaluate its main impacts on society.

    To simulate the reduction or increase of social distancing/quarantine, we remove/include some layers of the network, or change their edge weights. Similarly to the approach on \cite{ferguson2020report} to improve home contact when in quarantine, we increase the home layer edge weights by $20\%$ for each removed layer. To balance that we considered a smaller number of hours of contact in the base calculation for the home layer (3 hours a day), also taking into consideration that this layer has full contact between people of the same family. When the home contacts are increased according to our approach of layer removal, the time/intensity of contacts may increase up to its double.

\section{Results}

For each experiment with the proposed model, we consider the average and standard deviation (error) of 100 random repetitions to extract statistics of infection, death, and hospitalization time. Due to the random nature of these networks, it is possible that extreme cases occur within the repetitions, i.e. when the infection starts at a node that is not capable of further propagation, leading the epidemic to end at few iterations. Considering the real data we know that this is not the case, at least not for Brazil, therefore we manually remove these networks and they are not considered for the average/error calculations. It is important to notice, however, that this rarely happens, in all our experiments we noticed a maximum of 4 networks of this kind. Due to time and hardware constraints, our simulation considers 100,000 nodes, and the results need to be scaled up by a factor of 57 to match the Brazilian population statistics. This factor was empirically found by approximating the model results in the number of reported cases in Brazil. It is important to stress that for better statistics it should be considered the largest possible number of nodes to represent a population, i.e. the ideal case would be $n= total\ country/city\ population$. However, the computational cost of the simulation grows directly proportional to the number of nodes and edges of the network, and considering the critical situation of the moment at hand, 100,000 nodes are our limit to promptly present results of the epidemic dynamics.

In the experiments when varying the social distancing, the same network is considered in each iteration, i.e. comparisons of including/excluding layers are made in the same random network. We considered the epidemic began on February 26, which is the day the first confirmed case was officially reported. It is important to emphasize that we made various optimistic assumptions throughout the model construction and simulation, such as to consider that people are behaving with more caution by reducing direct contact, wearing masks, and doing proper home/hospital isolation when infected. It is also important to notice that we are not considering the number of available ICU/regular hospitalization beds for the death count, i.e. all the critical and severe cases are effectively treated. It is not trivial to estimate the direct impact of these numbers on the epidemic, however, this is an essential factor that directly impacts the number of deaths. Here we focus on the impacts of different actions on the overall epidemic picture, such as the increase and reduction of cases, deaths, and occupied beds in hospitals.

The social network starts normally, with all its layers and the original infection probabilities. The infection starts at a node with the closest degree to the average network degree and propagates at iterations of 1 day (up to 300 days). We consider an optimistic scenario, in which people are aware of the virus since the beginning, thus the initial infection probability is $\beta=0.3$. This represents a natural social distancing, a reduction of direct contacts that could cause infection (hugs, kisses, and handshakes), and also precautions when sneezing, coughing, etc. We empirically found that this initial value of $\beta$ yields results with a higher correlation to the Brazilian pandemic. A moderated quarantine is applied after 27 days, representing the isolation measures applied on March 24 by most Brazilian states, such as São Paulo \cite{SPquarentena}. To simulate this quarantine we remove the layers of religious activities and schools and reduce the contacts on transports and work down to $30\%$ of its initial value, i.e. $\beta=0.09$. The remaining activities on these layers represent services that could not be stopped, such as essential services, activities that are kept taking higher precautionary measures, and also those who disrespect the quarantine.

\subsection{Comparison to real data}

We compare the output of the model in the first 83 days with real data available from the Brazilian epidemic (up to May 18) \cite{EUODP,wometer,johnsHopkins,WHOcases}. The model achieves a significant overall similarity within its standard deviation. The greatest difference in the number of diagnosed cases at the last 10 days may be related to the increase in the number of tests being performed in Brazil, or yet, the constant decrease of isolation levels in the country (below $50\%$ for most days of the past month) \cite{mapaIsolamento}. We considered here a fixed isolation level around what was observed in the first days after the government decrees in Brazil, but data in ref. shows that these levels are constantly changing. Therefore, the number of diagnosed cases and deaths for the remaining simulation may be greater than the reported on this paper (see the "keep isolation" scenario in the next section).

\begin{figure}[!htb]
    \centering
    
    \subfloat[Daily new cases.]{ \includegraphics[width=0.48 \linewidth]{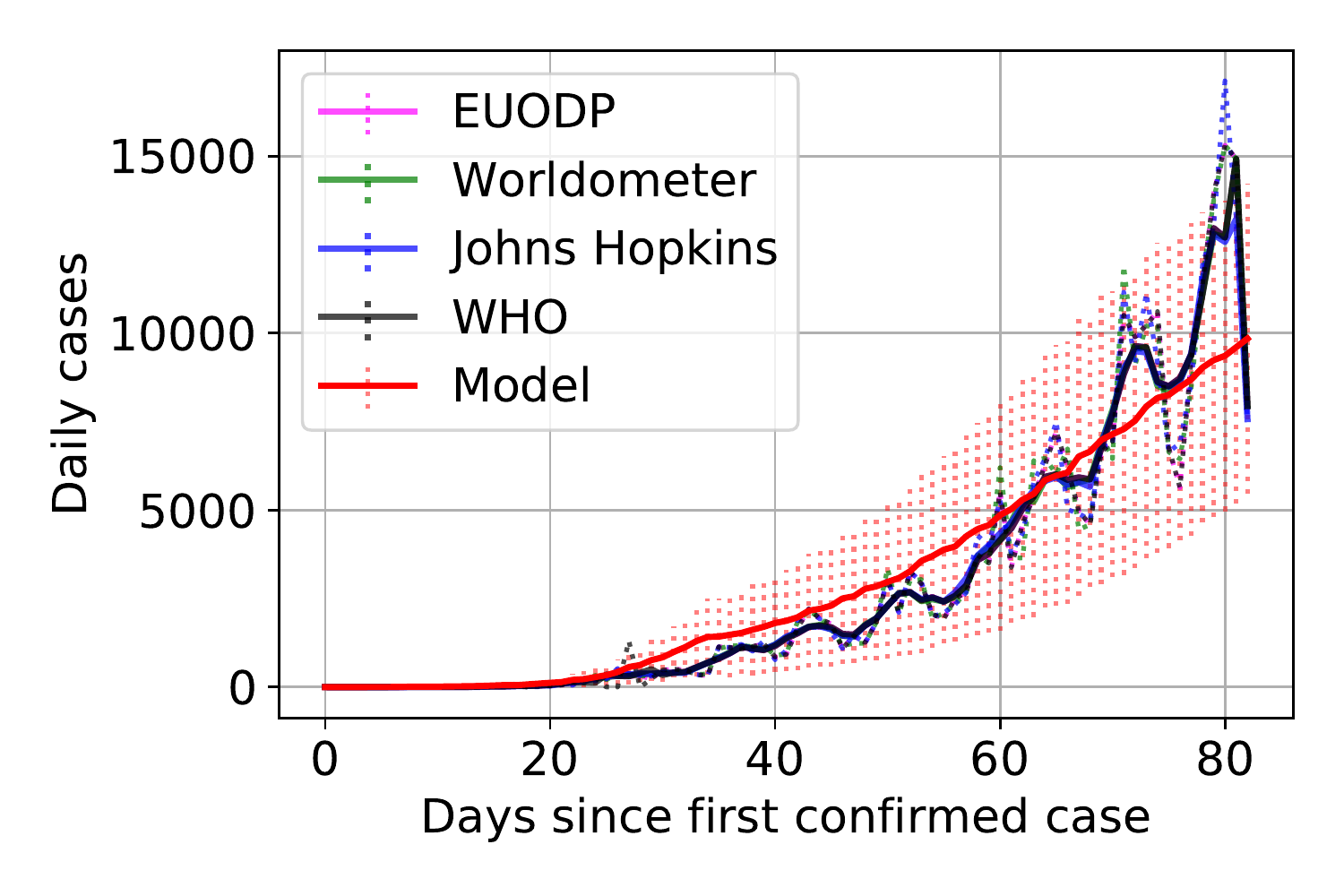}}  \subfloat[Daily new deaths.]{ \includegraphics[width=0.48 \linewidth]{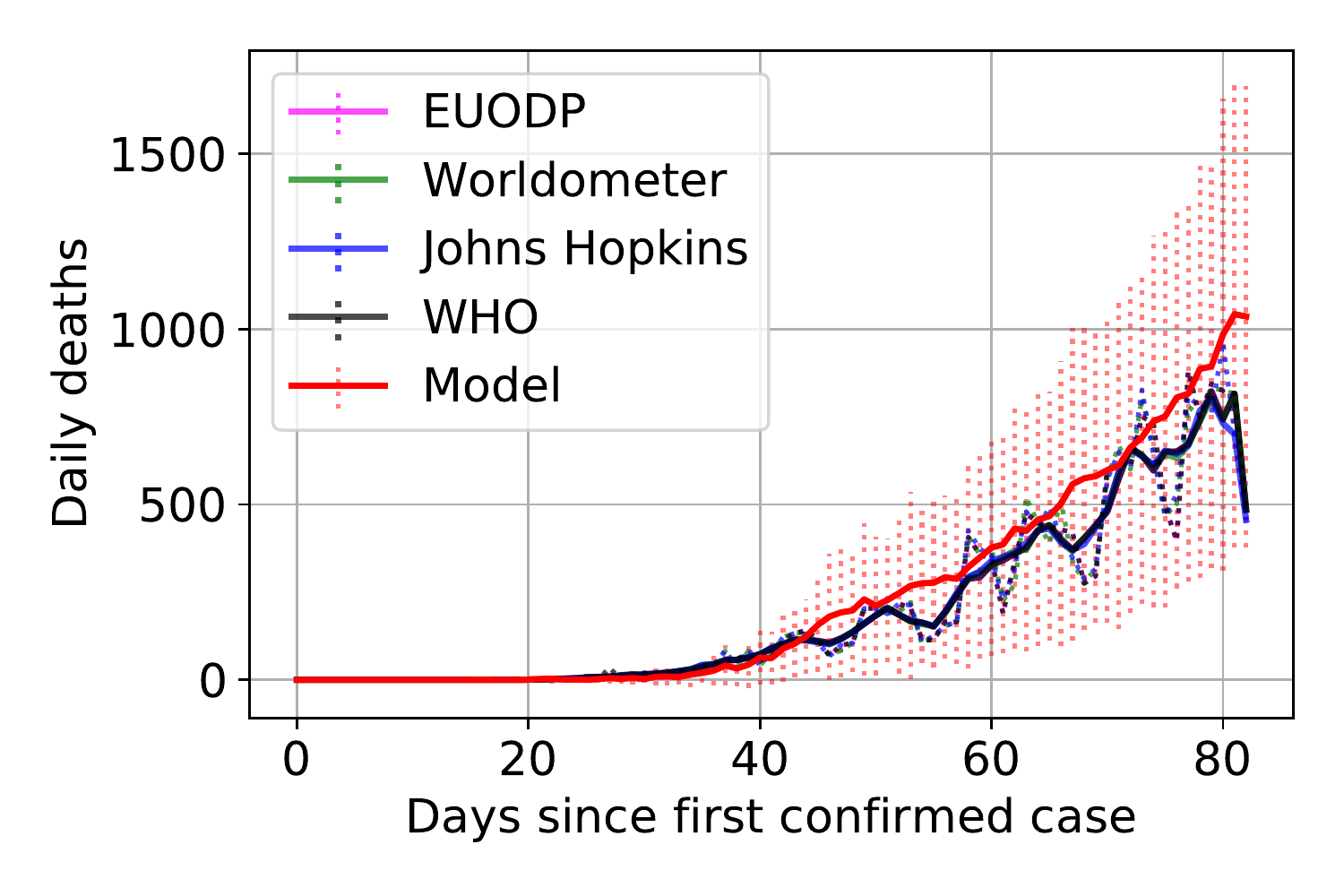}}
 
    \caption{\label{fig:comparison} Comparison between the proposed model output for the first 83 days (up to May 18) to 4 different data sources of the Brazilian COVID-19 numbers: EU Open Data Portal (EUODP) \cite{EUODP}, Worldometer \cite{wometer}, Johns Hopkins University \cite{johnsHopkins}, and World Health Organization (WHO) \cite{WHOcases}. The dotted lines represent the standard deviation, in the case of the real data the curve is the average over a 5-day window, and the solid lines the real raw data. The greatest average number of deaths produced by the proposed model may be related to underdetection (See Figure \ref{fig:subnot}).}

\end{figure}

\begin{figure}[!htb]
    \centering
    \includegraphics[width=0.72\linewidth]{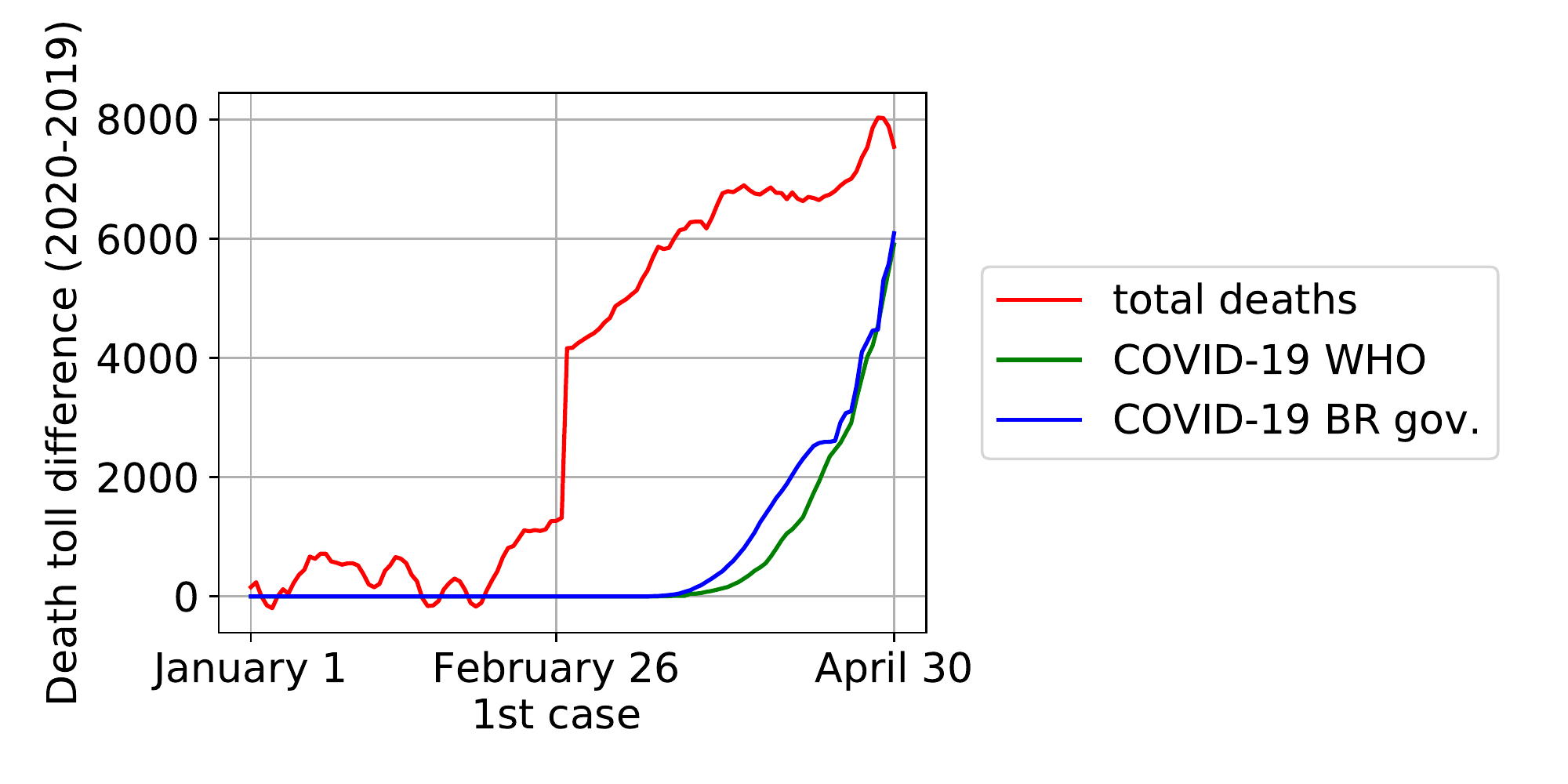}    
    \caption{\label{fig:subnot} To understand the impact of the COVID-19 underdetection in Brazil, we considered the official death records of 2019 and 2020 at the same period (January 1 to April 30) \cite{portalTransparencia}. Then the total death difference is compared to the COVID-19 records of the WHO \cite{WHOcases} and the Brazilian government \cite{portalTransparencia} data. The largest difference that appears right after the first confirmed case may indicate a significant underdetection of COVID-19 cases.}
\end{figure}

Concerning the daily death toll, the average number of the proposed model is greater than the official numbers. This is somehow expected, considering that the underdetection rates may be greater in contrast to the fewer number of tests being performed. To better understand this, we analyzed the number of death in Brazil from January 1 to April 30, comparing cases between 2019 and 2020, the results are shown in Figure \ref{fig:subnot}. It is possible to observe a clear increasing pattern after February 26, which is the day of the first officially confirmed case of COVID-19 in Brazil. This indicates that the real death toll for the disease may be significantly greater than the official numbers.

\subsection{Future actions and its impacts}

\begin{figure}[!htb]
    \centering
    \subfloat[Daily new cases (diagnosed)]{\includegraphics[width=0.5 \linewidth]{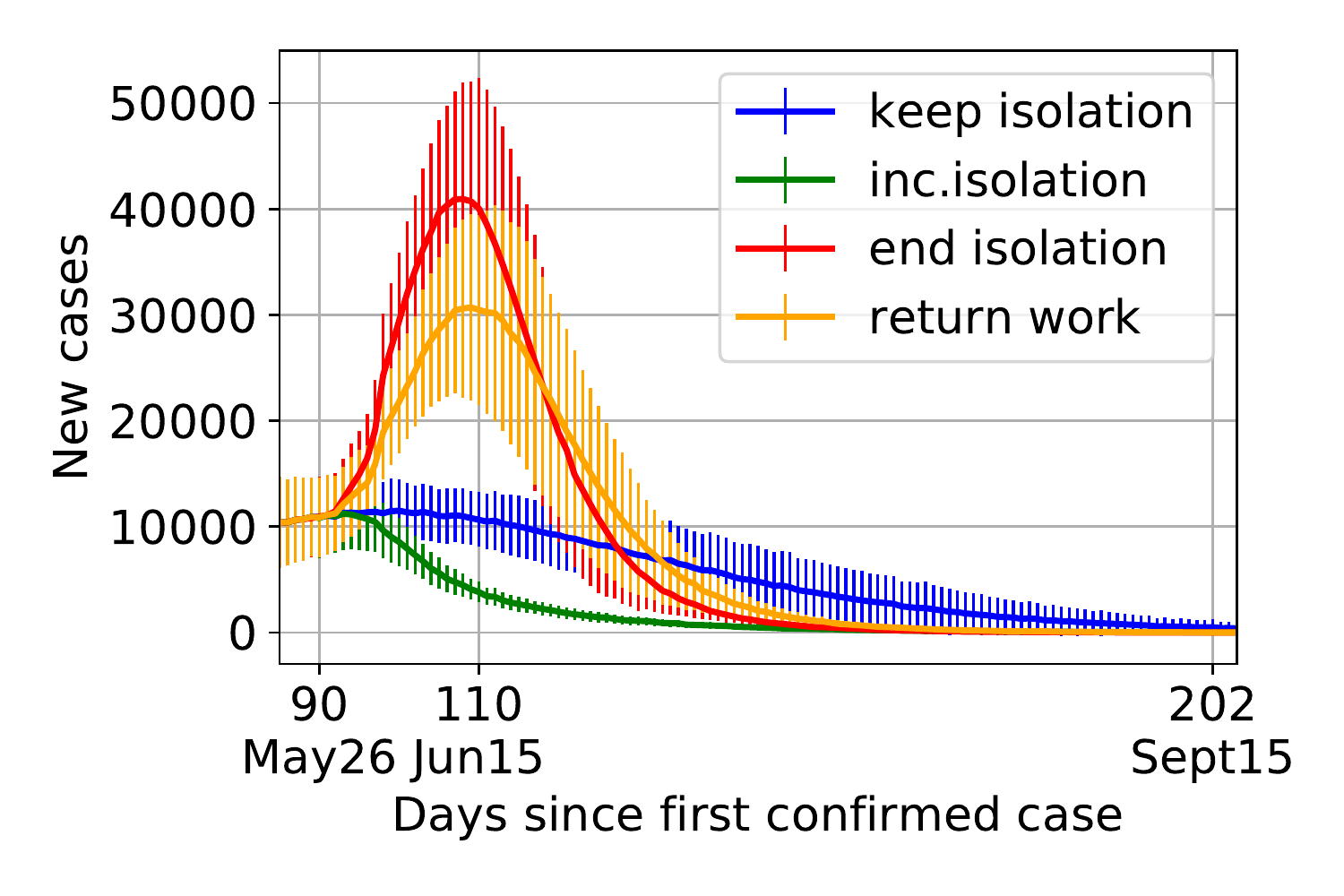}}\subfloat[Daily new deaths]{\includegraphics[width=0.5 \linewidth]{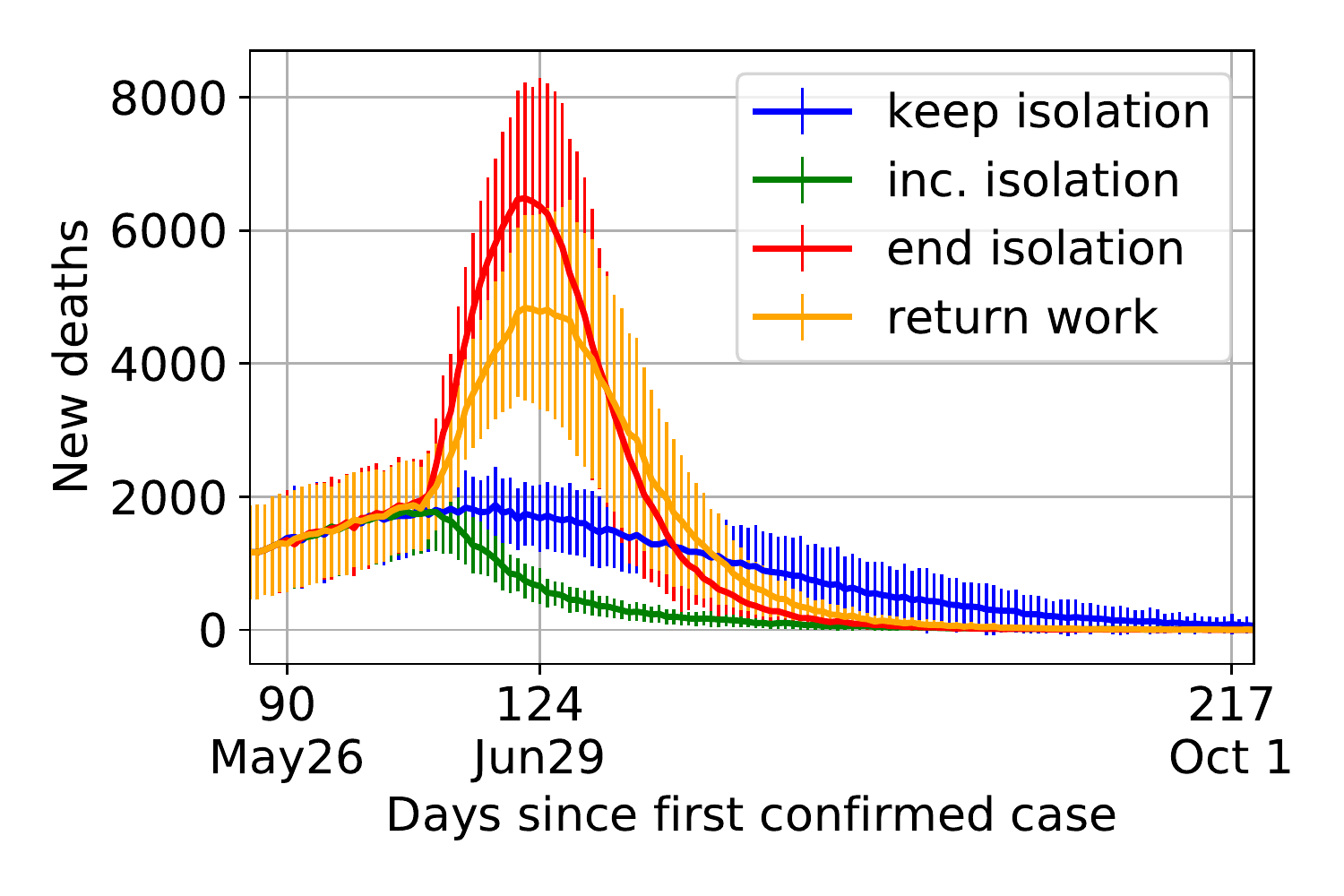}}
        
    \caption{\label{fig:exp_realistic_daily} Daily statistics in 4 possible scenarios after 90 days (May 26): Keep isolation levels; Increase isolation (stop work and public transports); End isolation (returns work and transport to normal and return school and religion); and return work (only the work layer is returned to normal).}
\end{figure}

After the initial epidemic phase, we consider 4 possible actions that can be taken after 90 days (May 26): a) Do nothing more, maintaining the current isolation levels; b) Stop isolation, returning activities to normal (initial network layers and weights); c) Return only work activities, restoring the initial probability of the layer; or d) Increase isolation, stopping the remaining activities in the work and transports layers (home and random remains). Firstly, we analyze the impacts on the number of daily new cases and deaths, results are shown in Figure \ref{fig:exp_realistic_daily}. As previously mentioned, at the start of the COVID-19 pandemic, Brazil was performing a fewer number of tests by an order of magnitude, in comparison to other countries with similar epidemic numbers, therefore we considered as diagnosed only the severe and critical cases, which are pronounced subjects for testing, and $20\%$ of the mild cases. The total infection ratio is discussed later. Considering keeping the current isolation levels, the peak of daily new cases occurs around 100 days after the first case (June 5), with around 11,000 confirmed cases. After 202 days (September 15), the average daily cases is around 500, and it goes below 100 daily cases after around 237 days (October 19). The peak of daily new deaths occurs around 118 days (June 23), with an average of 1900 deaths, and goes below 100 new occurrences after around 210 days (September 24). It is important to stress that this is a hypothetical scenario where the isolation level remains the same from day 27 to 300, which is hardly true in the real world where it is constantly changing \cite{mapaIsolamento}. The total numbers after the last day (300) account for 946,830 ($\pm 10,507$) diagnosed cases and 149,438 ($\pm 3,124$) deaths.

When we consider the return of all activities after 90 days, the number of cases and deaths grows significantly in an exponential fashion. The peak occurs at 108 days (June 13) with an average of 40,937 ($\pm$ 11,010) new cases, and at 122 days (June 27) with an average of 6,484 ($\pm$ 1,739) new deaths. Although the peak of cases/deaths and the decrease of the numbers occur early, in this case, the final result is critically worse, with a total of 1,340,367 ($\pm$ 18,513) diagnosed cases and 212,105 ($\pm$ 4,359) deaths. Here it is important to notice that we considered that all the activities return after 90 days and remain fully operational until the last day (300). Moreover, we do not account for the overloading of hospitals, which directly impacts the final death count. Therefore, the number of deaths may be considerably higher. Another possible scenario is the return of only the work layer, keeping reduced transports and no schools and religious activities, however, the pattern is similar to returning all activities, considering the growth time, peak, and decay time. The final numbers in this case are 1,253,119 ($\pm$ 26,009) diagnosed cases and 197,756 ($\pm$ 5,693) deaths.

If the isolation is strictly increased after 90 days (lockdown), the infection and death counts drop significantly in comparison to the other approaches. Moreover, the recovering time is much faster, as daily new cases stop earlier than the other scenarios. The peak of daily new cases happens around day 93 (June 1), and of daily new deaths around day 106 (June 11). The total numbers of diagnosed cases and deaths after day 300 are, respectively, 552,855 ($\pm$ 195,802) and 87,059 ($\pm$ 30,871).

\begin{figure}[!htb]
    \centering
    
    \subfloat[Keep isolation level.]{ \includegraphics[width=0.45 \linewidth]{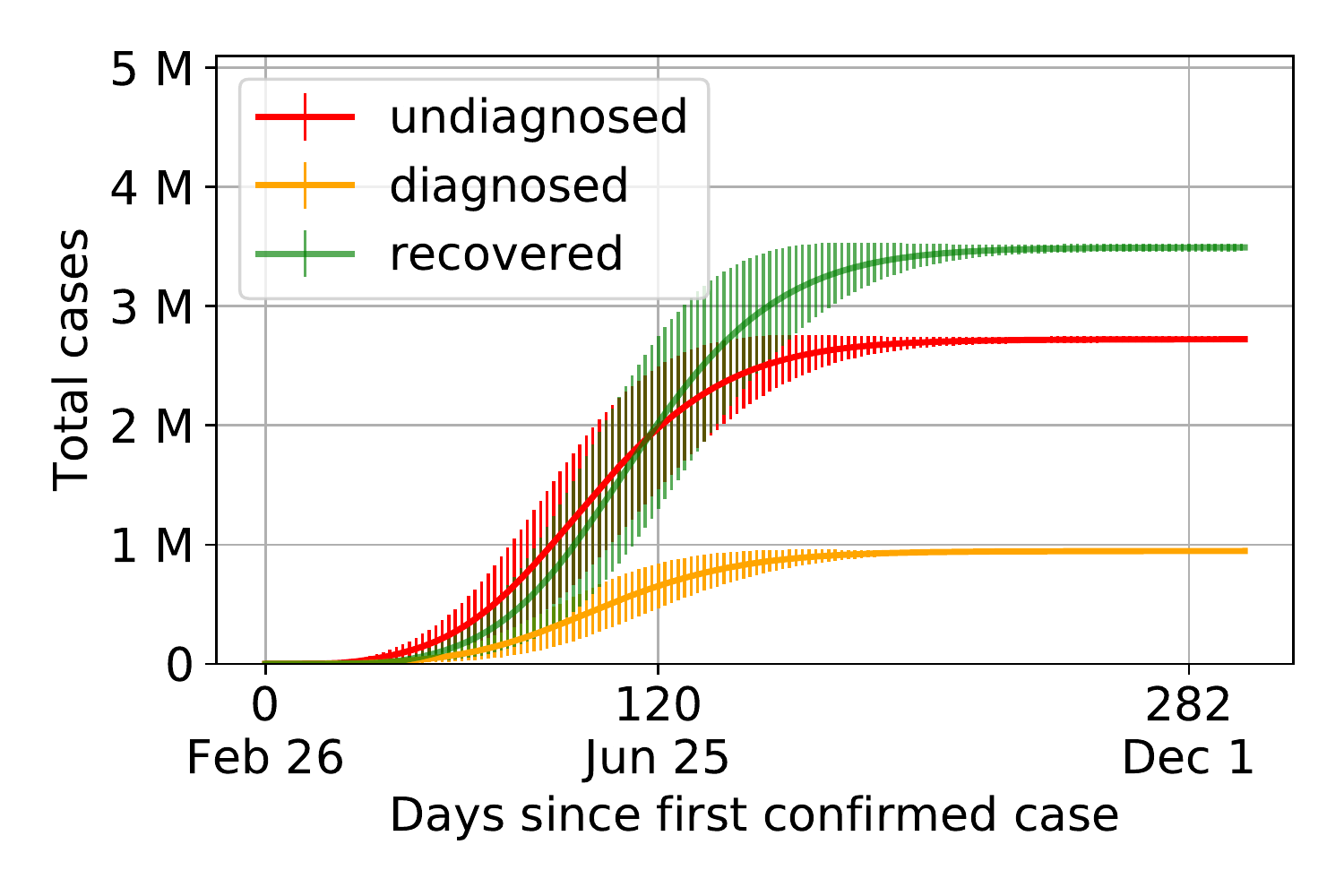} \ \ \includegraphics[width=0.45 \linewidth]{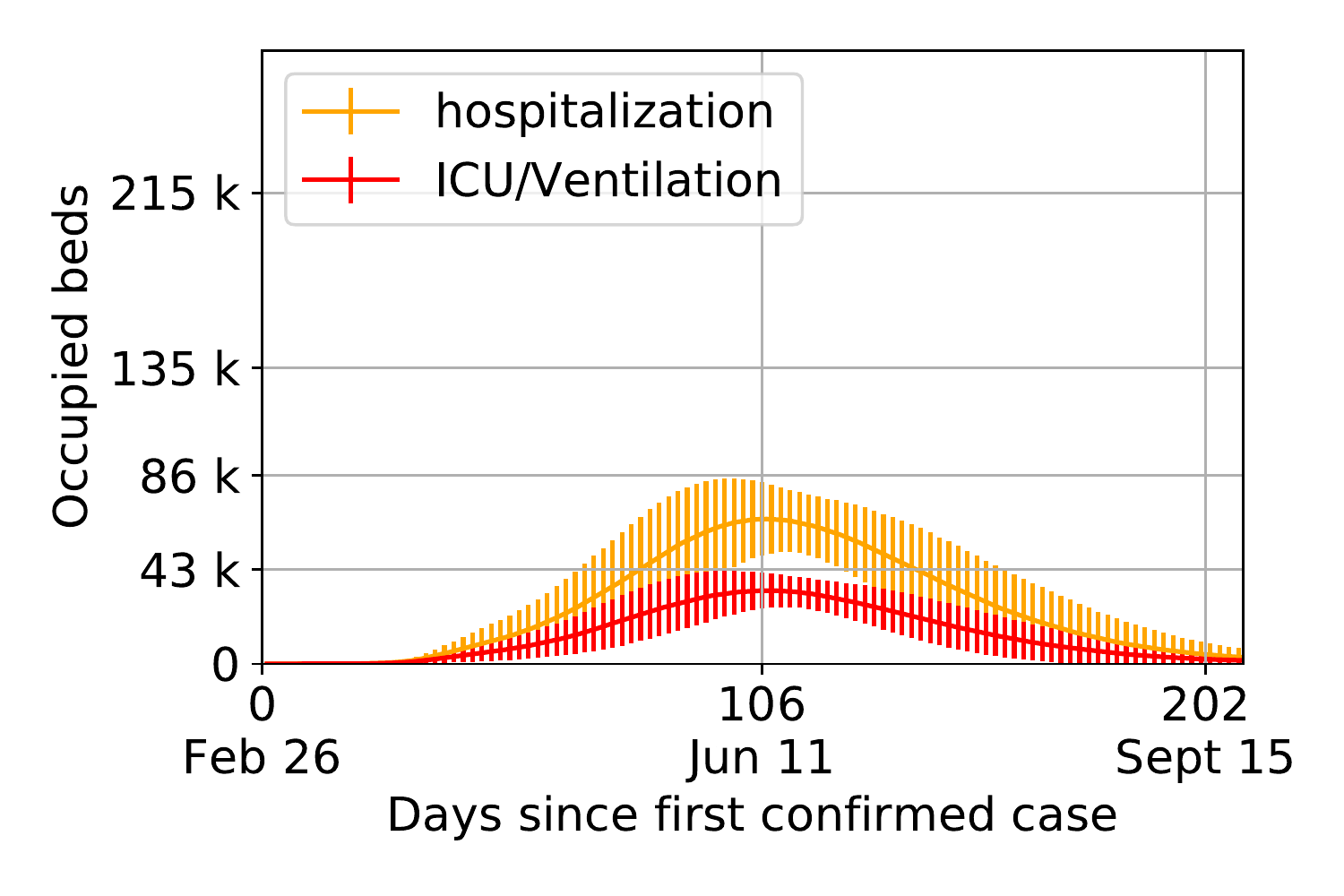}}

    \subfloat[End isolation, return all activities.]{\includegraphics[width=0.45 \linewidth]{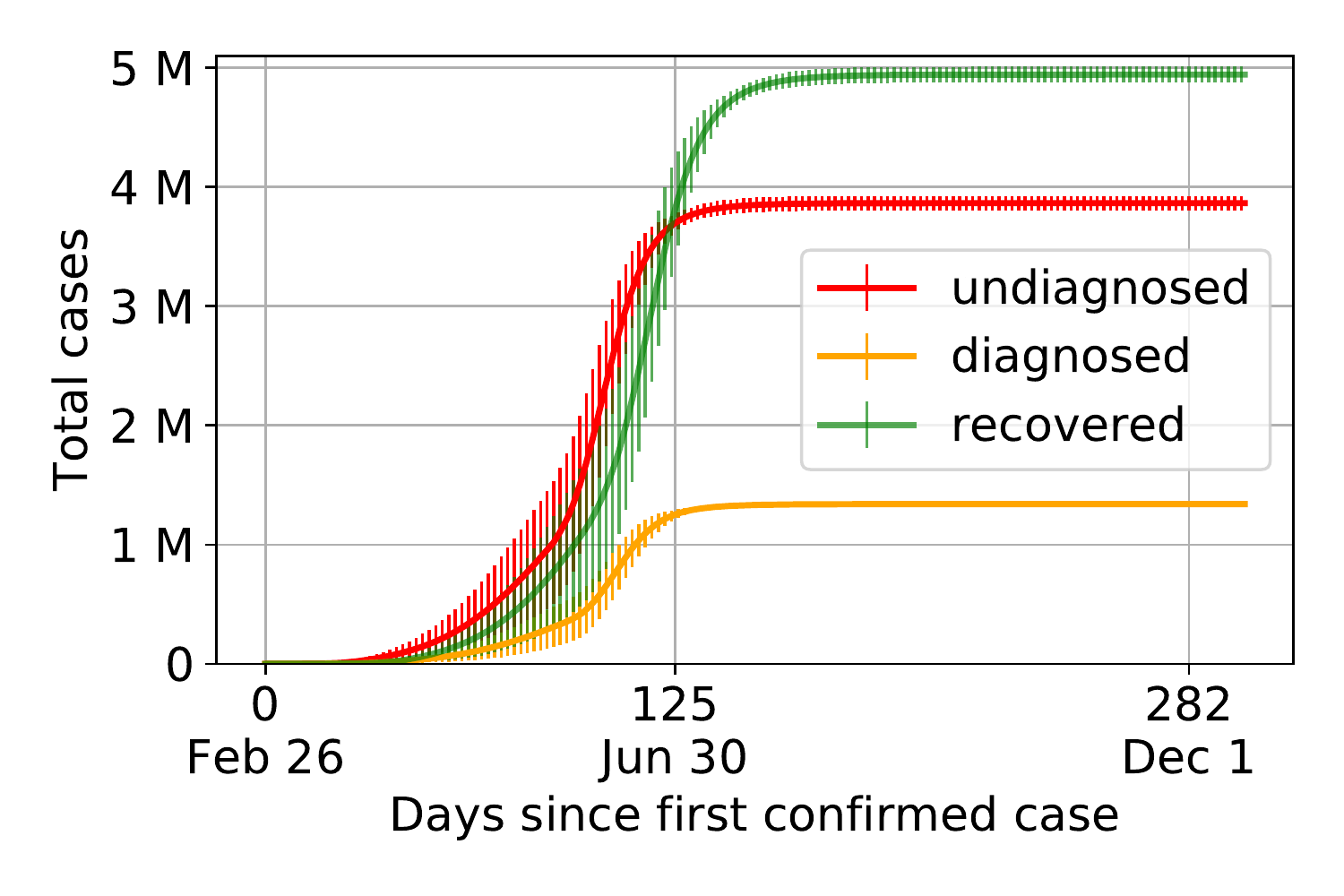} \ \  \includegraphics[width=0.45 \linewidth]{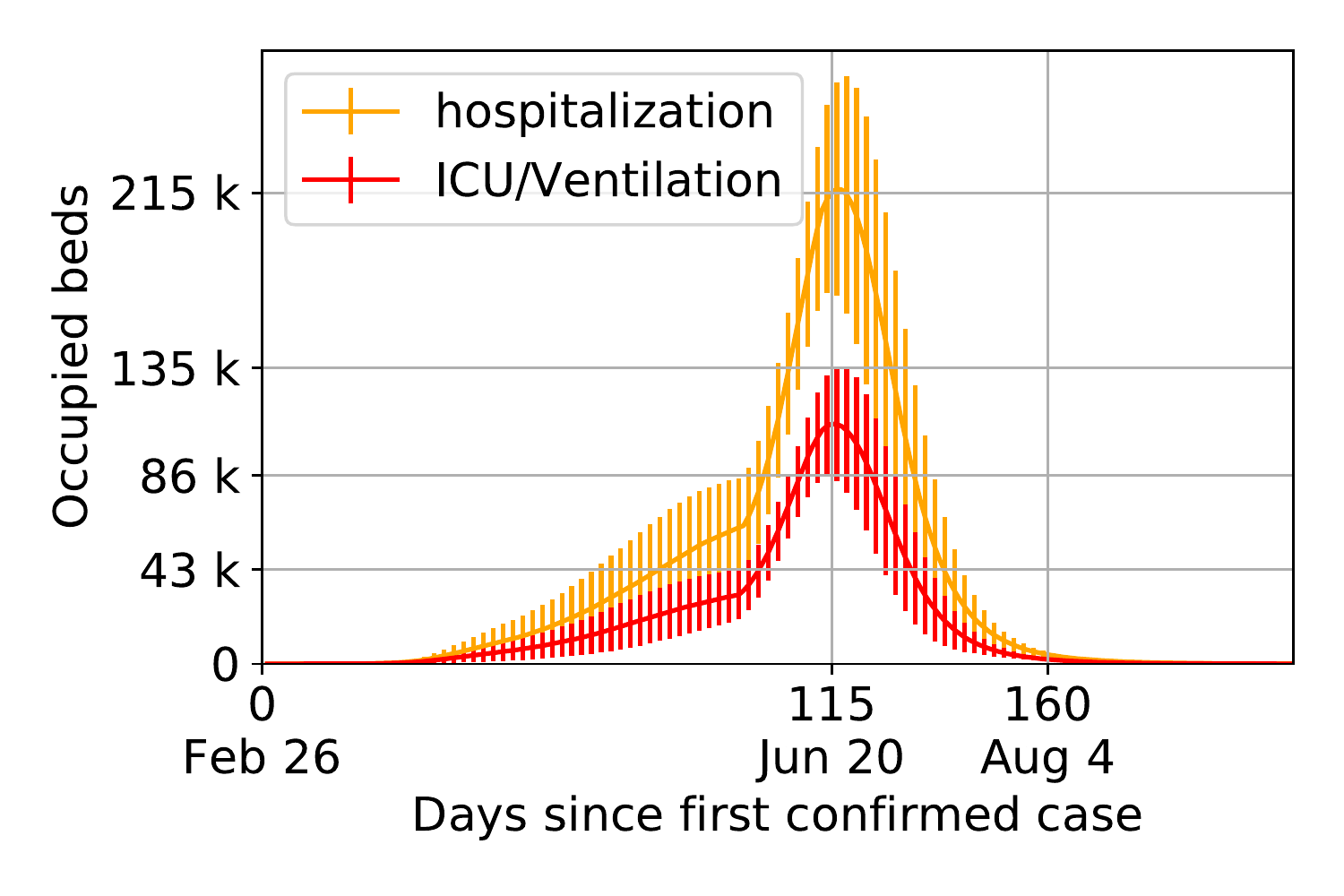}}

 \subfloat[Increase isolation.]{\includegraphics[width=0.45 \linewidth]{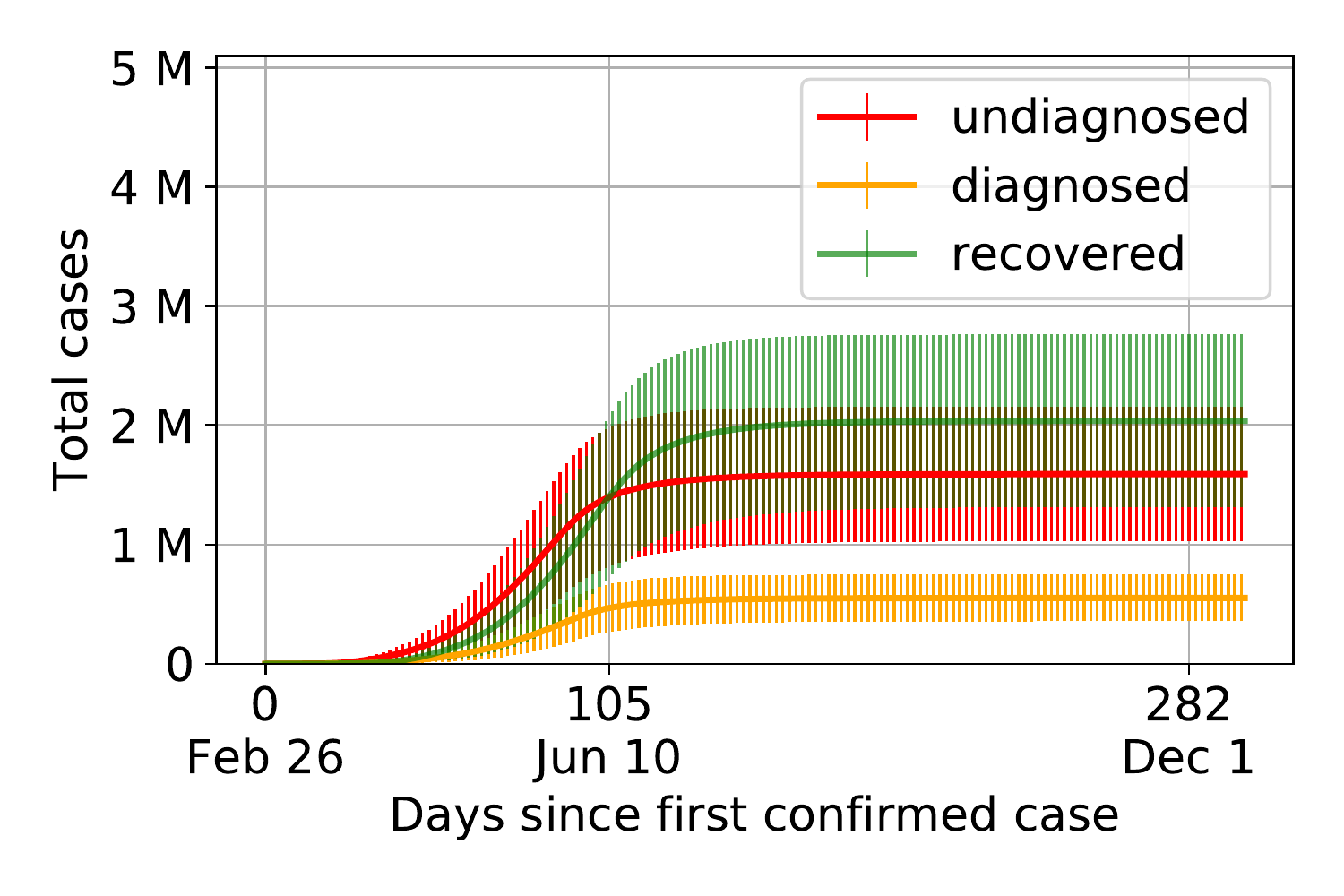} \ \  \includegraphics[width=0.45 \linewidth]{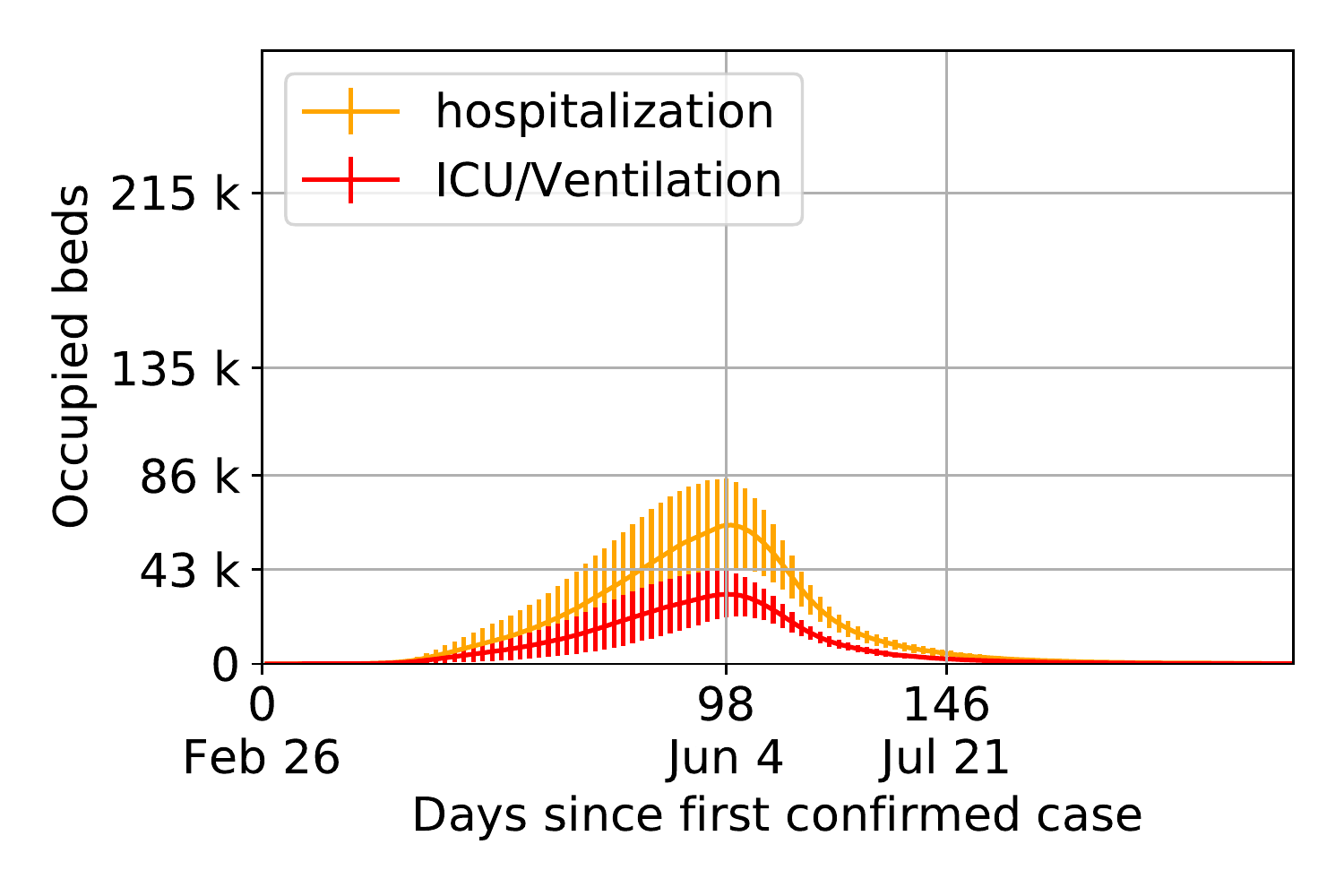}}
        
    \caption{\label{fig:exp_realistic_global} Total number of infected and recovered cases and evolution of hospital beds utilization in 3 possible scenarios after 90 days (May 26): (a) Keep isolation level (no schools and religion, reduced work and transports), (b) End isolation (return schools, religion, work and transport to normal) or (c) Increase isolation (stop work and public transports)).}
    
\end{figure}

Considering the hospitalization time described in the scheme of Figure \ref{fig:dynamic_configuration} it is possible to estimate the number of occupied beds for regular hospitalization (severe cases) and ICU/Ventilation (critical cases). We also show the difference between the cumulative growth of diagnosed and undiagnosed cases and recovered cases. The same approach as the previous experiment is considered (except for "return work") with 3 possible actions after 90 days (May 26), results are shown in Figure \ref{fig:exp_realistic_global}. The overall pattern of results is similar to the previously observed for the number of diagnosed cases and deaths. It is possible to notice that the number of undiagnosed cases is much higher than the diagnosed cases. This reflects the number of asymptomatic cases and the lack of tests for mild cases. In the worst scenario, which means ending the isolation, the total infected number may go above 5 million cases. The recovered rate is directly proportional to the infected rate, as one needs to be infected to either die or become resistant to the disease. If the infected rate is high, so is the recovered rate, e.g. the scenarios of keeping or ending isolation, and a high recovered rate also helps in mitigating the epidemic propagation (natural immunization). However, increasing isolation decreases the propagation much faster than natural immunization, with a considerably smaller death toll. It is also possible to observe the differences at the start of effective recovering, i.e. when the recovered rate surpasses the infected rates, this is due to the early increase in isolation levels.

The peak of hospitalization occupancy occurs around a week before the death peaks, in any scenario. In this case, ICUs are very important because critical patients are treated there, which represents the cases of higher death rates. Within the "end isolation" setting, patients may occupy up to an average of 215,285 ($\pm$ 48,682) regular beds and
109,520 ($\pm$ 24,647) ICU beds. These numbers are by far greater than entire Brazil's capacity, as publicly-available and private ICU beds sum up to 45,848 \cite{AMIB}. Even considering the better scenario, i.e. the lower bound of the standard deviation, the number of occupied ICU beds may reach around 86,000, which is also critical for Brazil's capacity (almost 2 times it's capacity). In this setting of "end isolation", the healthcare system would surely collapse.

When the isolation levels are kept, the numbers are significantly lower. However, the occupancy of 66,110 ($\pm$ 16,759) regular beds and 33,470 ($\pm$ 7,926) ICU beds is still critical for the Brazilian health system. Considering the creation of new provisional ICU units and good patient logistics, the situation may still remain under control during the peak of hospitalization occupancy. However, the results show that the hospital occupancy is prolonged considerably in this scenario, and they may stay functioning around their maximum capacity for up to a month (with an average of occupied ICU beds above 30,000). When increasing the isolation the peak of occupied beds is smaller, with an average of 63,226 ($\pm$ 20,682) regular beds and 31,816 ($\pm$ 10,592) ICU beds. Moreover, the shape of the curve throughout the days is different and the final numbers are considerably smaller. The peak also occurs around a week earlier and then decreases much faster. This scenario would be preferable as it has much more chances of not overloading the Brazilian healthcare system, relieving the hospital occupancy considerably faster and, therefore, contributing to the reduction of the number of deaths.

\section{Conclusion}    
    
This work presents a new approach for the modeling of the COVID-19 epidemic dynamics based on multi-layer complex networks. Each node represents a person, and edges are social interactions divided into 6 layers: home, work, transports, schools, religions, and random relations. Each layer has its own characteristics based on how people usually interact in that activity. The propagation is performed using an agent-based technique, a modification of the SIR model, where weights represent the infection probability that varies depending on the layers and the groups the node interacts, scaled by a $\beta$ term that controls the chances of infection. The network structure is built based on demographic statistics of a given country, region, or city, and the propagation simulation is performed at time iterations, that represent days. Here, we studied in depth the case of the Brazilian epidemic considering its population properties and also specific events, such as when the first isolation measures were taken, and the impacts of future actions.

Brazil is a large and populated country with a wide variety of geographical location types, climates, and it also has a lengthy border with other countries to the west. It is a challenging setting for any epidemiological study. Here we consider an average over all the country population, as we adjust the model output to match some statistics of the epidemic official reports. Brazil is performing fewer tests in comparison to other countries at the same epidemic scale, however, it is known that testing for infection is always limited, either due to the low number of tests or to the velocity of infections which the testing procedure cannot keep up to. We then considered that only hospitalization cases and $20\%$ of the mild cases are diagnosed. Asymptomatic cases are not diagnosed and keep acting normally in the network, considering the active layers. Regarding the isolation of infected nodes, we take some optimistic assumptions: Mild cases (even those not diagnosed) are aware of its symptoms and isolate themselves at home. Severe and critical cases are eventually hospitalized, and then fully isolated from the network (removal of all its edges).

Under the described scenario, the network starts with all its layers and $\beta=0.3$, representing that people are aware of the virus since the beginning (even before isolation measures). After 27 days of the first confirmed case, the first isolation measures are taken where schools and religious activities are stopped and work and transports keep functioning at $30\%$ of the initial scale (achieved further reducing the $\beta$ term). Different actions are then considered after 90 days of the first case: keep the current isolation levels, increase isolation, end isolation returning all activities to $100\%$, or returning only the work activities. The results show that keeping approximately the current isolation levels results in a prolonged propagation, as we are near the estimated peak (around June 5) with an average of 11,000 daily new cases and 1900 daily new deaths, and an average of 946,830 diagnosed cases (up to 3,6 million infected) and 149,438 deaths until the end of the year. In this scenario, hospitals may exceed its maximum capacity around June 11, but the efficient implementation of new ICU beds and good logistic management of patients may still keep the situation under control. However, this is a very optimistic assumption, considering that our definition of "keep isolation" considers social isolation above $50\%$ as registered at the beginning of the Brazilian quarantine \cite{mapaIsolamento}. The social isolation levels in Brazil are constantly decreasing even when we are still in a state of moderated quarantine, and it is possible to observe average isolation below $50\%$ in most days of the past month (middle of April to middle of May 2020). Moreover, the results show that this prolonged scenario may cause hospitals to keep functioning at maximum capacity for up to a month. When analyzing other possible scenarios the situation may be considerably different. Relaxing isolation measures from now on causes an abrupt increase in the daily growth of cases and deaths, up to 5 times higher in comparison to the current isolation levels. Even if only work activities return while schools, religion, and transport activities remain inactive/reduced, the impact is very similar to returning all the activities, with a possible number of above 1,34 million diagnosed cases (up to 5,2 million infected), and around 212,105 deaths until the end of the year. This is, again, a very optimistic assumption as we do not consider the hospital overflow to calculate the death toll. Considering this aspect, ICU beds may be fully occupied in early June, and around the middle of the month their demand may reach up to 134,000 beds, which is around 3 times higher than the entire country's capacity. The other alternative, which is the increase of isolation levels (lockdown), appears to be the only alternative to stop the healthcare system from entering a very critical situation. In this scenario, the growth in the number of daily cases and deaths would be mitigated, and faster. As we are near the peak of new cases at current isolation levels, estimated to be between the beginning and middle of June, increasing the isolation levels does not cause a significant impact on when the peak occurs or its magnitude. However, the disease spreading and the occurrences of new cases decrease much faster in this scenario in comparison to any other scenario studied here, with a difference of months. Moreover, the final numbers are considerably smaller, with an average of 552,855 diagnosed cases (up to 2.1 million infected) 87,059 deaths until the end of the year.

Although the proposed method includes various demographic information for the network construction, and an improved SIR approach to COVID-19, it still does not cover all factors that impact the epidemic propagation. As future works, one may consider more information such as the correlation between the age distribution within the social organization and the clinical spectrum of the 4 infection types (e.g. severe and critical cases are mostly composed of risk groups). Another possible improvement consists of increasing $n$ (number of nodes of the networks), e.g. using a value near the real population of the studied society, which we avoided here due to hardware and time constraints (graph processing is costly). Another important point regarding the obtained results is related to the "keep isolation" scenario, which may be underestimated as we take various optimistic assumptions and also consider a fixed isolation level based on previously observed data, while most recent data shows that these levels are decreasing \cite{mapaIsolamento}. Therefore, during the network evolution, a possible improvement is the use of dynamic isolation levels to better represent reality. It is also possible to consider various scenarios for future actions, such as 2 or more measures of increasing/reducing isolation. This may allow the discovering of new epidemic waves if social activities return too soon after the isolation period, such as what happened in 1918 with the Spanish flu.

\section*{Acknowledgments}

This study was financed in part by the Coordenação de Aperfeiçoamento de Pessoal de Nível Superior - Brasil (CAPES) - Finance Code 001. Leonardo F. S. Scabini and Lucas C. Ribas acknowledge support from São Paulo Research Foundation (FAPESP) (Grant \#2019/07811-0 and \#2016/23763-8). Altamir G. B. Junior acknowledges support from CNPq (Grant \# 144323/2019-2). Alex J. F. Farf\'{a}n and Mariane B. Neiva acknowledges support from CAPES (Grant number \#PROEX-9527567/D and CAPES PROEX-9056169/D respectively). Odemir M. Bruno acknowledges support from CNPq (Grant \#307897/2018-4) and FAPESP (grant \#2014/08026-1 and 2016/18809-9). He is also grateful to Henrique Pott Junior and Francisco Fambrini, for the fruitful conversation regarding the pandemic. 
The authors are also grateful to the NVIDIA GPU Grant Program for the donation of the Titan Xp GPUs used in this research.

\bibliographystyle{acm}
\bibliography{main.bib}

\end{document}